\documentclass[12pt]{article}
\usepackage[
top = 2.5cm, 
bottom = 2.5cm,  
left = 2.55cm,  
right = 2.55cm]{geometry}

\usepackage{bbm}
\usepackage[normalem]{ulem}
\usepackage{overpic}
\usepackage{subfigure}
\usepackage{caption}
\usepackage{mathtools}
\usepackage{latexsym} 
\usepackage{verbatim}  
\usepackage{tikz}
\usetikzlibrary{matrix}
\usepackage{color}
\usepackage{graphicx,amssymb,amsfonts,amsmath,amssymb,amscd,amstext, mathrsfs}
\usepackage{graphicx}
\usepackage{dsfont}
\usepackage{xcolor}
\usepackage{amsmath}
\usepackage{empheq}

\numberwithin{equation}{section}

\newlength\dlf

\def\CT{\mathcal{T}}

\newcommand{\bw}{\begin{widetext}}
\newcommand{\ew}{\end{widetext}}
\newcommand{\bea}{\begin{eqnarray}}
\newcommand{\eea}{\end{eqnarray}}
\newcommand{\be}{\begin{equation}}
\newcommand{\ee}{\end{equation}}

\renewcommand{\bar}[1]{\overline{#1}}
\renewcommand{\tilde}[1]{\widetilde{#1}}

\newcommand{\<}{\langle}
\renewcommand{\>}{\rangle}

\renewcommand{\cal}{\mathcal}
\newcommand{\CO}{\mathcal{O}}

\DeclareFontShape{OT1}{cmr}{mx}{n}{<->cmr10}{}
\newcommand{\titlefont}{\fontseries{mx}\selectfont}

\def\frac#1#2{{#1\over #2}}

\usepackage{jheppub}
\begin{document}

\begin{titlepage}

\begin{flushright} 
\end{flushright}

\begin{center} 

\vspace{0.35cm}

{\fontsize{20.5pt}{25pt}
{\titlefont 
LSZ in Action: Extracting Form Factors from Correlators Nonperturbatively in 2d $\phi^4$ Theory
}}

\vspace{1.6cm}  

{{A. Liam Fitzpatrick$^1$, Zhengxian Mei$^1$}}

\vspace{1cm} 

{{\it
$^1$Department of Physics, Boston University, 
Boston, MA  02215, USA
}}\\
\end{center}
\vspace{1.5cm} 

\begin{center} {\bf Abstract} \end{center}

{\noindent 
 In this paper, we  compute multiparticle form factors of local operators in 2d $\phi^4$ theory using a recently proposed method \cite{Henning:2022xlj} for efficiently implementing the LSZ prescription with Hamiltonian Truncation methods, and we adopt Lightcone Conformal Truncation (LCT) in particular for our calculations.   
We perform various checks of our results at weak and strong coupling, and elucidate the parametric behavior of truncation errors. 
 This opens up the possibility to compute S-matrix in various strongly coupled models using the LSZ method in LCT.
}

\end{titlepage}

\tableofcontents

%%%%%%%%%%%%%%%%%%%%%%
%%%%%%%%%%%%%%%%%%%%%%
\newpage

\section{Introduction and Summary} 

The S-matrix is probably the most important observable in all of particle physics, but computing it at strong coupling poses several challenges.  Often these challenges arise from the fact that the S-matrix is defined in terms of asymptotic states which strictly speaking require particles to be prepared in infinite volume, whereas most nonperturbative methods rely on finite volume.  Locality implies that it should be possible to approximately extract the S-matrix scattering amplitudes given a sufficiently large volume, turning the problem into a practical issue of how quickly a given finite-volume approximation approaches the infinite volume limit.\footnote{A well-developed approach on the lattice is L\"uscher's method for extracting elastic scattering amplitudes from the volume dependence of the finite-volume energy spectrum \cite{Luscher:1985dn,Luscher:1990ux}.}

 One nonperturbative methods that does at least formally allow infinite volume is Hamiltonian Truncation\footnote{See \cite{James:2017cpc} for a recent overview of Hamiltonian truncation techniques more generally, and \cite{Gabai:2019ryw} for a recent application of L\"uscher's method in Hamiltonian truncation. } evaluated in Lightcone Quantization, which exactly preserves a continuous set of boosts.  However, even here there is a sense in which finite volume effects sneak in. The problem is that in order to numerically diagonalize the Hamiltonian, one restricts the space of states to a finite-dimensional subspace of the full Hilbert space, and this subspace usually does not contain particles at arbitrarily large separation.  In more technical terms, this problem shows up in the following way.  Given the full set of time-ordered correlators of local operators of a theory, scattering amplitudes can be extracted using the LSZ prescription, by taking the residues of poles that arise when the external momenta are taken on-shell.  However, poles arising from the presence of multiple particles in the `in' state or `out' state' typically do not arise at finite truncation, but instead are essentially smeared out by truncation effects.

 Recently,  \cite{Henning:2022xlj} proposed a clever prescription for circumventing this problem. The idea was to extract the LSZ on-shell poles by taking matrix elements of derivatives of operators of the form
 \begin{equation}
 A(x) \equiv -(\partial^2 + m_p^2) \phi(x),
 \end{equation}
 where $m_p$ is the physical pole mass and $\phi$ is a local operator that creates the particle of interest, and then to use the equations of motion to express $A$ in terms of operators without time derivatives.  For instance, in $\phi^4$ theory, i.e., the theory of a real scalar field $\phi$ with a potential $V(\phi) = m_0^2 \phi^2/2 + \lambda \phi^4/24$, the equations of motion allow one to replace $A$ with
 \begin{equation}
 A(x) \cong \frac{m_0^2-m_p^2}{2} \phi + \frac{\lambda}{6} \phi^3,
 \label{eq:Aeom}
 \end{equation}
 as long as this expression is used inside Wightman correlation functions. This method was tested in the 3d $O(N)$ model at large $N$ in \cite{Henning:2022xlj}, as a method for computing scattering amplitudes.  Our goal in this paper will be to apply the same method in 2d $\phi^4$ theory, without invoking any large $N$ limit.  We will focus on a similar but comparably simpler observable, namely the two-particle form factor for the stress tensor $T_{\mu\nu}$ between the vacuum and a two-particle states:
 \begin{equation}
F(s) = \frac{1}{2} \langle \Omega | T^\mu_\mu(0) | p_1, p_2 \rangle.
\end{equation}
We will test our results by comparing them to perturbative calculations, specifically at one-loop and two-loops, as well as to strong coupling calculations from \cite{Chen:2021bmm} of this quantity computed using a different method.  We will compute all correlators using Lightcone Conformal Truncation (LCT), a lightcone quantization Hamiltonian Truncation method \cite{Katz:2016hxp,Anand:2020gnn}.  Much of the new technical work involved will be to efficiently compute the matrix elements of $\phi$ and $\phi^3$ between basis states when there is non-zero momentum exchange flowing through these operators.  We will present three separate methods for these quantities, which provides a strong consistency check on the calculations.  We also make example code available for implementations of all of them.

Probably the most important new qualitative feature that arises away from the large $N$ limit is that both terms $\phi$ and $\phi^3$ in (\ref{eq:Aeom}) contribute to the LSZ calculation.  This fact is less innocent than it might at first appear.  The point is that individually, the (Fourier transforms of ) correlators of $\phi$ and $\phi^3$ diverge on-shell, due to the presence of the LSZ pole $\sim 1/(p^2-m_p^2)$.  Therefore, it is not a priori obvious that in a numerical calculation, their sum will be sufficiently well-behaved to give even a finite answer, much less the correct one as the large truncation limit is taken.  Nevertheless, we will indeed see that in our perturbative calculations the divergent pieces in $\phi$ and $\phi^3$ do correctly cancel.  In fact, we will explicitly see that they individually grow like $\CO(\Delta_{\rm max})$ at large truncation, but with equal and opposite contributions to the correlator of $A$, and the residual finite pieces after this cancellation match the perturbative results from Feynman diagrams.  The convergence of the results is fastest at lower energies, and we will see the best results for the Taylor series expansion coefficients of the form factor as a function of $s$ in an expansion around $s=0$.  Larger values of $s$ converge more slowly, and in fact if we take the $s=\infty$ limit first then they never converge.  Turning to the strong coupling regime, a slightly more complicated picture emerges.  The divergent $\CO(\Delta_{\rm max})$ pieces in $\phi$ and $\phi^3$ still cancel, but at strong coupling there are $\CO(1/\Delta_{\rm max})$ truncation errors in the value for $m_0^2 - m_p^2$ in (\ref{eq:Aeom}) that contaminate the residual finite terms in the form factor.  We find that fixing the value of $m_0^2 -m_p^2$ by matching the form factor at a single value of $s$ leads to an accurate result over a very broad range of $s$.

This paper is organized as follows.  In section \ref{sec:LSZ}, we discuss the LSZ prescription in the context of Lightcone Conformal Truncation, and review the proposal from \cite{Henning:2022xlj}.  In section \ref{sec:MEmethods}, we describe three different methods for computing the new matrix elements we need in order to evaluate the form factor. In section \ref{sec:Feynman}, we compare our results in perturbation theory to the one- and two-loop results computed using Feynman diagrams.  In section \ref{sec:strong}, we compare our results at strong coupling to results from \cite{Chen:2021bmm}. Finally, in section \ref{sec:future}, we conclude with a discussion of potential future directions.

\section{Application of LSZ in truncation}
\label{sec:LSZ}

\subsection{Applying LSZ to two-particle form factor}

Our main object of interest is the two-particle form factor for an operator $\CO$, which can be extracted by applying LSZ to the following correlator:
\begin{equation}
G_{\CO}(p_1, p_2) \equiv \int d^2 y d^2 z e^{-i (p_1 \cdot y + p_2 \cdot z)} \< \Omega | \CT \{ \CO(0) \phi(y) \phi(z) \} | \Omega\>.
\label{eq:Gdef}
\end{equation}
We will work in lightcone coordinates for the momenta, $p_{\pm} = \frac{p_{0} \mp p_{1}}{\sqrt{2}}$, $p^2 = 2 p_+ p_-$.  There are two cases we need to consider, one where $p_{1-}$ and $p_{2-}$ have the same sign, and one where they have different sign; without loss of generality, by Hermitian conjugation we can take $p_{1-}>0$.    Because there are no physical states with $p_-<0$, we have
\begin{equation}
\begin{array}{cc} \tilde{\phi}(p) | \Omega\> = 0 & \textrm{ if } p_-<0, \\  
\< \Omega | \tilde{\phi}(p) = 0 & \textrm{ if } p_->0, \end{array}  \qquad  \tilde{\phi}(p) \equiv \int d^2 x e^{-i p \cdot x} \phi(x).
\end{equation}

 Take the case where both $p_{1-}$ and $p_{2-}$ are positive.  Because of the time-ordering and the observation above, contributions to the integral vanish unless $y^+ <0$ and $z^+<0$. We can focus on the case $0> y^+>z^+$ and obtain the case $0>z_+ >y_+$ by symmetrizing $p_1 \leftrightarrow p_2$. Moreover, we will insert a complete set of states as follows:
 \begin{equation}
 \mathbbm{1} = \sum_i \int_0^\infty \frac{d p_-}{(2\pi) 2p_-} | \mu_i, p_-\> \< \mu_i, p_- |,
 \end{equation}
 where $\mu_i$ labels the (in our truncated system, discrete) set of eigenvalues of the operator $P_+$, through $P_+ | \mu_i, p\> = \frac{\mu_i^2}{2p_-} | \mu_i, p_-\>$, and $p_-$ labels the (continuous) set of eigenvalues of the operator $P_-$.  
  So we have
 \begin{equation}
 \begin{aligned}
 G_\CO(p_1, p_2) &= G_\CO^+ (p_1, p_2) + G_\CO^+(p_2, p_1), \\ 
 G^+_\CO(p_1, p_2) &= \int_{0> y^+ > z^+} d^2 y d^2 z e^{-i (p_1 \cdot y + p_2 \cdot z)} \\
 & \sum_{i,j} \int \frac{dp_- dp_-'}{(2\pi)^2 2p_- 2p_-'}\< \Omega | \CO(0) |\mu_i, p_-\> \< \mu_i, p_-| \phi(y) | \mu_j, p'_-\> \< \mu_j, p'_-| \phi(z) | \Omega\> .
 \label{eq:GO}
\end{aligned}
\end{equation}
Now,  since $\phi(x) = e^{i P \cdot x} \phi(0) e^{-i P \cdot x}$, we can perform the integrals over $y,z,p$ and $p'$ to obtain
\begin{equation}
G_\CO^+(p_1, p_2) = 
(i)^2  \sum_{i,j}  \frac{ \< \Omega | \CO(0) | \mu_i,(p_1+p_2)_-\>\< \mu_i, (p_1+p_2)_- |\phi(0) | \mu_j, p_{2-} \> \< \mu_j, p_{2-} | \phi(0) | \Omega\>
 }{ (p_2^2 - \mu_j^2+i \epsilon)(2(p_{1}+p_2)_- (p_{1+} + \frac{\mu_j^2}{2p_{2-}}) - \mu_i^2+i \epsilon)} .
 \end{equation}
To create a two-particle asymptotic state, LSZ instructs use to  read off the residue of the pole at $p_1^2=m_p^2$ and $p_2^2 = m_p^2$.  The pole at $p_2^2=m_p^2$ is manifest because the single-particle state is an eigenstate of the Hamiltonian, i.e., it is just the eigenstate with $\mu_j^2=m_p^2$. We define the usual wavefunction overlap factor $Z$ as
\begin{equation}
\< \Omega | \phi(0) | p_2\> \equiv Z^{1/2} .
\end{equation}
Reading off the residue of this pole and dividing by $Z^{1/2}$, we obtain
\begin{equation}
\begin{aligned}
F^+_\CO(p_{1-}, p_{1+}, p_{2-}) &\equiv -i \frac{1}{\sqrt{Z}}\textrm{Res}_{p_{2+} = \frac{m_p^2}{2p_{2-}}} G_\CO^+(p_1, p_2) \\
 &= i \sum_i \frac{\< \Omega | \CO(0) | \mu_i , (p_1+p_2)_-\>\< \mu_i, (p_1+p_2)_- | \phi(0) | p_{2-}\>}{(p_1+p_2)^2 - \mu_i^2+i \epsilon} \Big|_{p_{2+} = \frac{m_p^2}{2 p_{2-}}} ,
\end{aligned}
\end{equation}
where $|p_{2-}\>$ refers to the one-particle state with lightcone momentum $p_{2-}$.

The function $F_\CO^+$ depends on three variables, but in our lightcone truncation setup there is an exact invariance under boosts which  we can use to  boost to a momentum frame where $(p_1+p_2)_-=1$.  We will do so implicitly from now on unless explicitly stated otherwise, and to reduce clutter we will denote $|\mu_i, (p_1+p_2)_-\>$ as $|\mu_i\>$.  That leaves only two variables, which we will generally parameterize in terms of $p_{2-}$ and $s$ defined as
\begin{equation}
s \equiv (p_1 + p_2)^2.
\end{equation}
In terms of these variables, we have 
\begin{equation} 
F_\CO^+(s, p_{2-}) = i \sum_i \frac{\< \Omega | \CO(0) | \mu_i \> \<\mu_i | \phi(0) | p_{2-} \>}{s- \mu_i^2 + i \epsilon}.
\label{eq:PreFF}
\end{equation}
There is a manifest pole in this expression when $s$ passes through the value of any physical state, as we expect.  However, what is more difficult to see is where the pole at $p_1^2=m_p^2$ comes from.  At fixed $s$, all dependence on $p_1^2$ enters through its implicit dependence on $p_{2-}$, which appears in the matrix elements $\< \mu_i | \phi(0) | p_{2-}\>$ in the numerator.  To make this dependence on $p_1^2$ explicit, we can solve the following relation for $p_{2-}$ in terms of $s$ and $p_1^2$:
\begin{equation}
s =2 (p_{1-} + p_{2-})(p_{1+} + p_{2+}) =  \frac{p_1^2}{1-p_{2-}} + \frac{m_p^2}{p_{2-}} .
\end{equation}
 The ambiguity in the choice of which solution one takes to this equation for $p_{2-}$  goes away after adding the symmetric combination $G^+(p_1, p_2) + G^+(p_2, p_1)$ from (\ref{eq:GO}). 

In the large truncation limit, the pole in $p_1^2$ has to emerge from the sum over the amplitudes $\< \mu_i | \phi(0) | p_{2-}\>$.  We will see that these functions are all individually polynomials in $p_{2-}$, so a true pole cannot emerge at any finite value of truncation.\footnote{We thank Matthew Walters for discussions on this point.} Instead, we will be forced to apply the LSZ prescription indirectly.

Before moving on to strategies to overcome this issue, let us briefly revisit the second case referred to above, where $p_{1-}>0$ and $p_{2-}<0$.  In this case, due to the time-ordering, the only integration region in $y^+$ and $z^+$ that contribute are $y^+<0 $ and $z^+>0$.  Inserting a complete set of states as before,
\begin{equation}
\begin{aligned}
G_\CO(p_1, p_2) & = \int_{z^+ > 0 > y^+} d^2 y d^2 z e^{-i (p_1 \cdot y+ p_2 \cdot z)} \\
& \sum_{ij} \int \frac{dp_- dp'_-}{(2\pi)^2 2p_- 2p'_-} \< \Omega | \phi(z) | \mu_i, p_-\>\< \mu_i , p_-| \CO(0) | \mu_j, p'_-\> \<\mu_j, p'_- | \phi(y)  | \Omega\>.
\end{aligned}
\end{equation}
The analysis proceeds similarly to before, except that now both the poles in $p_1$ and $p_2$ arise from one-particle eigenstates and are manifestly visible even at finite truncation.  Taking $p_1^2=p_2^2=m_p^2$ and reading off the  residue of the pole from the one-particle eigenstates, one obtains
\begin{equation}
- \textrm{Res}_{p_1^2=m_p^2, p_2^2=m_p^2} \frac{1}{Z} G_\CO(p_1, p_2) = \< p_1 | \CO(0) | p_2 \>.
\label{eq:TChannel}
\end{equation}
This quantity can therefore be easily computed in the truncation framework, as was done in \cite{Chen:2021bmm}.  Its main disadvantage is that it requires taking one in state and one out state, and so does not directly probe the interesting kinematic region with a two-particle asymptotic state.

\subsection{Introducing Equation of Motion}\label{eom}

In the previous subsection, we discussed why it is difficult in truncation to directly read off the residue of poles in order to create multi-particle asymptotic states. The strategy proposed in \cite{Henning:2022xlj} is to evaluate the differential operator $i(\partial^2 + m_p^2)$ acting on $\phi$ before taking the momenta on-shell.  That is,  we compute
\begin{equation}
\tilde{G}^+_\CO(p_1, p_2) \equiv \int d^2 y  e^{-i p_1 \cdot y}  i(\partial_y^2 + m_p^2) \< \Omega | \CT \{ \CO(0) \phi(y)  \} | p_2\>.
\end{equation}
It is crucial that the mass $m_p$ is the physical mass, not the bare mass.  Note that we have already used the manifest pole for the first particle to 
produce $|p_2\>$ in the ket state.  When the derivatives $\partial_y^2$ hit the time-ordered correlator, there is a contribution from the time-derivative hitting the time-ordering $\theta$ functions, and another contribution where the derivatives all act within the time-ordering brackets.  Consider the former contribution. Writing $\partial^2 =2 \partial_+ \partial_-$ and the time-ordering as 
\begin{equation}
\<\Omega| {\cal T} \{ \CO(0) \phi(y)\} | p_2 \> \equiv \theta(-y^+) \< \Omega | \CO (0) \phi(y) | p_2 \> +\theta(y^+) \< \Omega |  \phi(y) \CO(0) | p_2 \>,
\end{equation}
we have
\begin{equation}
2 \partial_+ \partial_-  \<\Omega |  {\cal T} \{ \CO(0) \phi(y)\} | p_2 \>  =2 \delta(y^+) \<\Omega |   [\partial_- \phi(y),  \CO(0)] | p_2 \> +  \<\Omega |  {\cal T} \{ \CO(0) \partial^2\phi(y)\} | p_2 \> .
\end{equation}
The first term on the RHS can typically be simplified by using the canonical commutation relations or, if $\CO$ is a conserved current, Ward identities.  We will mainly focus on the case where $\CO= T_{--}$ is the energy-momentum tensor, in which case the commutator can be evaluated for instance as follows.  
 In lightcone quantization, the canonical commutation relations are
\begin{equation}
[ \phi(x), \partial_- \phi(y) ] = \frac{i}{2} \delta(x_- - y_-),
\end{equation}
and therefore
\begin{equation}
[ \partial_- \phi(y) , T_{--}(x)] =
   i \partial_- \phi(x) \frac{\partial}{\partial y^-} \delta(y_- - x_-),
 \end{equation}
 since $T_{--} = (\partial \phi)^2$.  Substituting this into the expression for $\tilde{G}_{\CO}^+$, it generates the following contribution:
 \begin{equation}
\tilde{G}_{\CO}^+ \supset \int d^2 y e^{-i p_1 \cdot y} 2i \delta(y^+) \< \Omega | [ \partial_- \phi(y), T_{--}(0) ]| p_2\> =-2 p_{1-} p_{2-},
\end{equation}
 where we have used $\< \phi(x) | p\> = e^{ -i p \cdot x}$.  This term is the tree-level contribution, which we will separate out from the rest of the form factor and concentrate on computing the remaining term, from $ \<\Omega |  {\cal T} \{ \CO(0) \partial^2\phi(y)\} | p_2 \>$.  When $p_{-1} = -p_{2_-}$, (i.e., $s=0$), the tree-level contribution does not receive any higher order corrections, due to the Ward identity for $T_{\mu\nu}$. 

The remaining terms, where $\partial^2+m_p^2$ act within the time-ordering, are more complicated.  Inserting a complete set of states and repeating the analysis that led to (\ref{eq:PreFF}), the only change is that the terms where $i(\partial^2+m_p^2)$ acts inside the time-ordering become
\begin{equation}
\tilde{F}_\CO^+(s,p_{2-}) \equiv  \sum_i \frac{\< \Omega | \CO(0) | \mu_i\> \< \mu_i | (\partial^2 + m_p^2) \phi(0) | p_{2-}\>}{s-\mu_i^2 + i \epsilon}.
\label{eq:Ftilde}
\end{equation}
The strategy proposed in \cite{Henning:2022xlj} is that we can evaluate $\partial^2\phi$ directly, using the equations of motion in $\frac{\lambda}{4!} \phi^4$ theory:
\begin{equation}
\partial^2  \phi =-m_0^2 \phi -\frac{\lambda}{6} \phi^3.
\end{equation}
Note that $m_0$ in this equation is the bare Lagrangian mass, not the physical mass $m_p$.  

This strategy explicitly relies on using a Lagrangian theory, so that one knows the equations of motion for the fields.  Before we discuss how to implement this approach, we will first take a step back and describe a more general strategy that only relies on the Hamiltonian itself, which can be obtained in non-Lagrangian theories.\footnote{We thank Ami Katz for suggesting this method.}  The trick of using the equations of motion above will then emerge naturally as a special case.

To begin,  we write $\partial^2 \phi$ as $2P_+$ acting on $\partial_- \phi$, i.e., $\partial^2 \phi = 2i [P_+, \partial_- \phi ]$, in (\ref{eq:Ftilde}):
\begin{equation}
\tilde{F}_\CO^+(s,p_{2-}) = -  \sum_i \frac{\< \Omega | \CO(0) | \mu_i\> \< \mu_i | 2i  [P_+, \partial \phi(0)]+m_p^2 \phi(0) | p_{2-}\>}{s-\mu_i^2 + i \epsilon}.
\label{eq:Ftilde2}
\end{equation}
To evaluate the commutator, we can use the fact that we know how to compute the matrix elements of $P_+$ and $\partial \phi$ explicitly in a basis of our choosing, say with basis states $|k\>$:
\begin{equation}
\< \mu_i | 2  [P_+, \partial \phi(0)] | p_{2-}\> = 2 \sum_k  \< \mu_i | P_+| k \> \< k| \partial \phi(0) | p_{2-}\> - \< \mu_i | \partial \phi(0) | k\> \< k | P_+ | p_{2-}\>.
\label{eq:comm2}
\end{equation}
When we sum over the intermediate states $|k\>$, we have the option to insert a complete set of basis states $|k\>$ for a much larger subspace than the truncated basis we use when we diagonalize the Hamiltonian.  In fact, it is not too hard to see that it is necessary to take a larger basis for the intermediate states.\footnote{Here is how.  If the basis for the intermediate states $|k\>$ is the same as the basis we used to diagonalize the Hamiltonian, then we can equivalently write the sum over basis states $|k\>$ as the sum over eigenstates $|\mu_i\>$ of $P_+$.  Then the matrix elements of the commutator reduce to
\begin{equation}
\< \mu_i | 2  [P_+, \partial \phi(0)] | p_{2-}\> = \left( \mu_i^2 - \frac{m_p^2}{p_{2-}} \right) \< \mu_i |  \partial \phi(0) | p_{2-}\>,
\end{equation}
where we recall we are using the frame $p_{1-}+p_{2-}=1$. 
 In this case, (\ref{eq:Ftilde2}) becomes
\begin{equation}
\tilde{F}_\CO^+(s,p_{2-}) =  \sum_i  \left( \frac{m_p^2}{p_{2-}(1-p_{2-})} - \mu_i^2 \right) \frac{\< \Omega | \CO(0) | \mu_i\> \< \mu_i |  \partial \phi(0) | p_{2-}\>}{s-\mu_i^2 + i \epsilon}.
\label{eq:Ftilde3}
\end{equation}
Note that when both particles are on-shell, then $s=\frac{m_p^2}{p_{2-}(1-p_{2-})}$, so the term in parenthesis is just $s-\mu_i^2$. Therefore (\ref{eq:Ftilde3}) (incorrectly!) has no imaginary part, since $\textrm{Im}(\frac{x}{x+i \epsilon}) =-\pi x\ \delta(x) = 0$. }

In a Lagrangian theory, there is a slick way to evaluate this commutator with an infinite truncation for the intermediate basis states $|k\>$ in (\ref{eq:comm2}): we can use the canonical commutation relations.  This way of evaluating the commutator is equivalent to using the Euler-Lagrange equations of motion for $\phi$.  To see this explicitly, take the $P_+$ Hamiltonian for $\frac{\lambda}{4!} \phi^4$ theory:
\begin{equation}
P_+ = \int dx^- \frac{m_0^2}{2} \phi^2 + \frac{\lambda}{4!} \phi^4,
\end{equation}
and use the canonical commutation relations to evaluate $[P_+, \partial \phi]$:
\begin{equation}
[P_+, \partial \phi(y)] =  \int dx^- \frac{m_0^2}{2} [ \phi^2(x), \partial \phi(y)] + \frac{\lambda}{4!} [ \phi^4(x), \partial \phi(y)] = \frac{i}{2} \left( m_0^2 \phi(y) + \frac{\lambda}{6} \phi^3(y)\right). 
\end{equation}
Substituting into (\ref{eq:Ftilde2}), 
\begin{equation}
\tilde{F}_\CO^+(s,p_{2-}) =  \sum_i \frac{\< \Omega | \CO(0) | \mu_i\> \< \mu_i |  \delta m^2 \phi(0) + \frac{\lambda}{6} \phi^3(0) | p_{2-}\>}{s-\mu_i^2 + i \epsilon},
\label{eq:FtildeFinal}
\end{equation}
where we have defined the mass shift as
\begin{equation}
\delta m^2 \equiv m_0^2 - m_p^2.
\end{equation}
The advantage of this expression is that we have essentially cancelled the ``$\frac{p_1^2-m_p^2}{p_1^2-m_p^2}$'' factor
by using the equations of motion, so that we should be able to take both $p_1$ and $p_2$ momenta on-shell from the beginning.  Imposing this constraint, and adding in the symmetric combination $\tilde{F}_\CO^-$, we finally obtain  the following expression for the on-shell form factor:
\begin{equation}
\boxed{
\begin{aligned}
\tilde{F}_\CO(s) &=  
\tilde{F}_\CO^+(s,p_{2-,\rm os}) + \tilde{F}_\CO^+(s,1-p_{2-, \rm os}), \\
\tilde{F}_\CO^+(s,p_{2-}) &=  \sum_i \frac{\< \Omega | \CO(0) | \mu_i\> \< \mu_i |  \delta m^2 \phi(0) + \frac{\lambda}{6} \phi^3(0) | p_{2-}\>}{s-\mu_i^2 + i \epsilon},
\end{aligned}
}
\label{eq:FtildeFinal}
\end{equation}
where $p_{2-, \rm os}$ and $(1-p_{2-, \rm os})$  are the two on-shell values of $p_{2-}$, i.e., the solutions to $s=\frac{m_p^2}{p_{2-}(1-p_{2-})}$. 
In order to evaluate (\ref{eq:FtildeFinal}), we need to construct matrix elements for $\phi$ and $\phi^3$ between arbitrary basis states.  The new ingredient compared to previous work is that the bra state and ket state can have different $p_-$ momentum, so that there is some momentum flowing through the operators $\phi$ and $\phi^3$ themselves.\footnote{The work \cite{Chen:2021bmm} evaluated matrix elements of $T_{--}$ between basis states with different momenta. That case was simplified by the fact that $T_{--}$ is a primary operator, so the momentum dependence was completely fixed by conformal symmetry.}

\section{Computing $\phi$ and $\phi^3$ matrix elements}
\label{sec:MEmethods}

In this section, we describe three separate methods for calculating the matrix elements of $\phi$ and $\phi^3$ in a basis of states created by Fourier transforms of chiral primary operators acting on the vacuum:
\begin{equation}
| \CO_i, p_- \> \equiv \frac{1}{N_{\CO_i}} \int dx^- e^{i p_- x^-} \CO_i(x^-) | 0\>.
\label{eq:BasisStateRep}
\end{equation}
   Previous work \cite{Anand:2020gnn} developed efficient methods for calculating the matrix elements of the Hamiltonian $P_+$ in this basis, and the methods presented here will be generalizations that allow momentum to flow through $\phi$ and $\phi^3$.  
The first method works directly with the Fock space of the free massless scalar field.  All basis states are expressed in terms of their wavefunctions in the multiparticle Fock space, and matrix elements of $\phi$ and $\phi^3$ reduce to integrals over all the individual particle momenta.  We call this the  `Jacobi Polynomials' method because the wavefunctions of states (\ref{eq:BasisStateRep}) created by primary operators are products of Jacobi polynomials.  The second and third method instead first evaluate correlators of operators in position space, and obtain all the required matrix elements by performing Fourier transforms of these correlators; unlike the first method, only two integrals (one for the bra state and one for the ket state) have to be performed in this case.  The second (`Wick Contraction') method evaluates the position space correlators by performing Wick contractions, and does not use (or rely on) the conformal invariance of the free massive scalar theory.  By contrast, the third (`Radial Quantization') method significantly simplifies the evaluation of the position space correlators by using the fact that conformal transformations can be used to map the bra and ket primary operators to 0 and $\infty$, where they can be represented in terms of radial quantization modes.\footnote{  
These three methods, for matrix elements between bra and ket states with equal momentum, are described in more detail in \cite{Anand:2020gnn}.  Additionally, in the special case of matrix elements of $T_{--}$ between bra and ket states with differing momentum, the Radial Quantization method was described in \cite{Chen:2021bmm}, which was simplified by the fact that $T_{--}$ is itself a primary operator.  By contrast,  $\phi^3$ is not a primary operator, and computing its matrix elements is consequently more involved.}
The computational efficiency of the Radial Quantization method is significantly higher than the other two methods and we were able to produce results up to $\Delta_{\operatorname{max}}=40$. Having all three methods is a useful consistency check and we present numeric code for them at \href{https://github.com/andrewliamfitz/LCT}{https://github.com/andrewliamfitz/LCT}.  Below, we will briefly describe these approaches, focusing on the novel aspects required for generalizing previous techniques to the case of the $\phi$ and $\phi^3$ form factors. 

\subsection{Jacobi Polynomial Method}\label{jacobi}

The simplest representation of the basis states we use in our truncation is in terms of the multiparticle Fock space states.  Each single particle state is labeled by its $p_-$ momentum and has the following inner product norm:
\begin{equation}
\< p_- | p'_-\> = 2 \pi 2 p_- \delta(p_- - p'_-).
\end{equation}
Each basis state $|\CO, p_-\>$ is labeled by a primary operator $\CO$ and momentum $p_-$.  The free theory multi-particle form factors $F_{\CO}(p) \equiv \<p_{1-}, \dots, p_{n-}| \CO(0)\>$ determine the state $|\CO, p_-\>$ in terms of the massless theory Fock space states:
\begin{equation}
|\CO, p_-\> = \frac{1}{n! N_{\CO}} \int \frac{d p_{1-} \dots dp_{n-}}{(2\pi)^2n 2p_{1-} \dots 2p_{n-}} (2\pi) \delta(p_- - \sum_{i=1}^n p_{i-}) F_{\CO}(p_1, \dots, p_n)|p_{1-}, \dots, p_{n-}\>.
\label{eq:FockState}
\end{equation}
Here, the number of particles $n$ depends on the primary operator $\CO$.  
Because the operators $\CO$ are defined to be primaries, each wavefunction $F_\CO$ is an eigenfunction of the conformal Casimir,
\begin{equation}
  \mathcal{C}=-2\sum_{i<j} p_ip_j\left(\frac{\partial}{\partial p_i}-\frac{\partial}{\partial p_j}\right)^2,
\end{equation}
which implies that the $F_\CO$s are linear combinations of products of Jacobi polynomials of the following form \cite{Anand:2020gnn}:
\begin{equation}
  F_{\boldsymbol{l}}\equiv p_1...p_n |p|_n^{l_n} \prod_{i=1}^{n-1} |p|_{i+1}^{l_i}P_{l_i}^{(2|l|_{i-1}+2i-1,1)}\left(\frac{p_{i+1}-|p|_i}{|p|_{i+1}}\right),
\end{equation}
where $\boldsymbol{l}=(l_1,l_2,...l_n)$, $P$ is the Jacobi polynomial, and
 $ |p|_i\equiv \sum_{j=1}^i p_j$.
 In addition, because of Bose symmetry, the physical states are linear combinations of these polynomials that are symmetric in all the $p_i$s.  
Schematically, the corresponding primary operators are
\begin{equation}
  F_{{\boldsymbol{l}}}(p_i) \Leftrightarrow \mathcal{O}_{\boldsymbol{l}} \sim \partial^{l_n}\bigg(\partial \phi \overleftrightarrow{\partial}^{l_{n-1}}\bigg(\partial\phi...\overleftrightarrow{\partial}^{l_2}\bigg(\partial\phi\overleftrightarrow{\partial}^{l_1}\partial\phi\bigg)\bigg)\bigg).
\end{equation}
Given a representation of the form (\ref{eq:FockState}) for the bra and ket states, any matrix element of powers of $\phi$ can be computed by integrating over all the $p_i$ momenta.  Computing matrix elements of $\phi$ and $\phi^3$ between bra and ket states with differing momentum in this method is straightforward, and we provide example code that does this explicitly. 
 The main disadvantage is that the integrals become very computationally expensive very quickly as the number of particles in the external states grows.

\subsection{Wick Contraction}\label{wickcontraction}

Instead of using the wavefunctions directly at an early stage, another attempt is to keep the operator expression until the very end. The primary operators in our basis  can all be expressed as a linear combination of ``monomials'' \cite{Anand:2020gnn}
\begin{equation}
  \mathcal{O}=\sum_{\mathbf{k}} \mathcal{C}_{\mathbf{k}}^{\mathcal{O}} \partial^{\mathbf{k}}\phi,
\end{equation}
where $a_n$ are numbers, and $\partial^{\mathbf{k}_n}\phi$ are monomials. As an example, for $\mathbf{k}_n=(1,2,2)$, 
\begin{equation}
  \partial^{(1,2,2)}\phi\equiv\partial\phi\,\partial^2\phi\,\partial^2\phi .
\end{equation}

For any two-point function of primary operators, we can then expand it in the form of 
\begin{equation}
  \langle \mathcal{O}(x)\mathcal{O}'(z)\rangle = \sum_{\mathbf{k},\mathbf{k'}} \mathcal{C}_{\mathbf{k}}^{\mathcal{O}}\mathcal{C}_{\mathbf{k'}}^{\mathcal{O}'} \langle \partial^{\mathbf{k}}\phi(x)\partial^{\mathbf{k'}}\phi(z)\rangle
  \label{eqn:wickexp}
\end{equation}
Similarly, when we want to evaluate the three-point function of $\phi^3$, i.e. the $\phi^3$ matrix element in the orthoprimary basis, we can expand it as  
\begin{equation}
  \langle \mathcal{O}(x)\phi^3(0)\mathcal{O'}(z)\rangle = \sum_{\mathbf{k},\mathbf{k'}} \mathcal{C}_{\mathbf{k}}^{\mathcal{O}}\mathcal{C}_{\mathbf{k'}}^{\mathcal{O}'} \langle \partial^{\mathbf{k}}\phi(x)\phi^3(0)\partial^{\mathbf{k'}}\phi(z)\rangle
\end{equation}
Where we can always set $\phi^3$ to be at the origin due to translational invariance. With known coefficients for each composite monomial of a primary operator, the question is now to compute the three-point function of monomials. We can do this by performing Wick contraction, and rewrite the three-point function of monomials as a product of a three-point function and a two-point function, which we have closed form expressions for 
\begin{equation}
  \langle \partial^{\mathbf{k}}\phi(x)\phi^3(0)\partial^{\mathbf{k'}}\phi(z)\rangle=\langle \partial^{k_i}\phi(x)\phi^3(0)\partial^{k'_{j,l}}\phi(z)\rangle\times\langle\partial^{\mathbf{k}/k_i}\phi(x)\partial^{\mathbf{k'}/k'_{j,l}}\phi(z)\rangle
\end{equation}
\begin{equation}
  =\langle \partial^{k_i}\phi(x)\phi(0)\rangle\times\langle\phi(0)\partial^{k'_j}\phi(z)\rangle\times\langle\phi(0)\partial^{k'_l}\phi(z)\rangle\times\langle\partial^{\mathbf{k}/k_i}\phi(x)\partial^{\mathbf{k'}/k'_{j,l}}\phi(z)\rangle
\end{equation}
\begin{equation}
  =\frac{\Gamma(k_i)}{x^{k_i}}\frac{\Gamma(k'_j)}{(-z)^{k'_j}}\frac{\Gamma(k'_l)}{(-z)^{k'_l}}\times\frac{A_{\mathbf{k}/k_i,\mathbf{k'}/k'_{j,l}}}{(x-z)^{|\mathbf{k}/k_i|+|\mathbf{k'}/k'_{j,l}|}}
\end{equation}
where $\mathbf{k}/k_i$ means removing the element $k_i$ from vector $\mathbf{k}$, and 
\begin{equation}
  A_{\mathbf{k},\mathbf{k'}}\equiv\sum_{\mathbf{\sigma}\in\operatorname{perm}(\mathbf{k'})}\Gamma(k_1+\sigma_1)...\Gamma(k_n+\sigma_n)
\end{equation}
and recursively we can also write 
\begin{equation}
  A_{\mathbf{k},\mathbf{k'}}=\sum_{i=1}^n \Gamma(k_n+k'_i)A_{\mathbf{k}/k_n,\mathbf{k'}/k'_i} .
\end{equation}
Here we are contracting the $\phi^3$ in the middle with one $\phi$ on the left and two $\phi$'s on the right. Note that we require that $|\mathbf{k}/k_i|=|\mathbf{k'}/k'_{j,k}|$ after Wick contraction, therefore $\mathbf{k'}$ always has one more element than $\mathbf{k}$. This means that the number of particles in state created by $\mathcal{O'}$ is always one more than that by $\mathcal{O}$. We call this the $n\rightarrow n+1$ piece of $\phi^3$ matrix. Intuitively, this means that $\phi^3$ annihilated one particle on the left, and two particles on the right, so that the remaining has a non-zero overlap. 
Similarly, we can construct the $n\rightarrow n-1$, $n\rightarrow n+3$, and $n\rightarrow n-3$ pieces of $\phi^3$, as well as the $n\rightarrow n+1$ and $n\rightarrow n-1$ pieces of $\phi$. All of them have similar closed form expressions, therefore we can compute the exact three-point function of monomials in position space. 

The next step is to Fourier transform this expression into momentum space, so that we can use the LSZ prescription and Equation of Motion as presented in the previous section. Notice that all three-point functions of $\phi$ and $\phi^3$ we evaluate take the form of 
\begin{equation}
  \frac{\operatorname{const.}}{x^A(-z)^B(x-z)^C}
\end{equation}
Since the constant will not affect the result of Fourier transform, we can simply focus on 
\begin{equation}
  \int dx\,dx'\frac{e^{i(Px-P'z)}}{(x-i\epsilon)^A(-z-i\epsilon)^B(x-z-i\epsilon)C}
\end{equation}
which has an analytic solution 
\begin{equation}
  \frac{4\pi^2 P^{A+C-1}P'^{B-1}}{\Gamma(A+C)\Gamma(B)}\Theta(P)\Theta(P')_2F_1(C,1-B,A+C,P/P')
  \label{eq:FTAnalytic}
\end{equation}

Now we have an expression for three-point functions with monomials in momentum space, one simply needs to sum over these results with appropriate coefficients to obtain the three-point functions with primary operators, i.e. the matrix elements of $\phi$ and $\phi^3$.

\subsection{Radial Quantization}

In section \ref{jacobi}, we presented the Jacobi method which computed integrals of Jacobi polynomials in a brute-force way. This method works in any QFT, without restriction to CFTs. In section \ref{wickcontraction}, the Wick contraction method was also general for any QFT, but we used conformal symmetries to obtain a closed-form expression for the two-point and three-point functions. In this section, we present the radial quantization method, which is a way of quantizing states in a CFT, allowing us to exploit the orthogonality between states generated by monomials, giving much higher coputational efficiency.

In radial quantization, we expand $\partial\phi$ as 
\begin{equation}
  \partial\phi(x) = \frac{i}{\sqrt{4\pi}}\sum_{k=1}^\infty \sqrt{k}\bigg(x^{-k-1}a_k+x^{k-1}a_k^\dagger\bigg), \qquad
  [a_k,a_k'^\dagger]=\delta_{k,k'},
\end{equation}
where $a_k$, $a_k^\dagger$ are the annihilation and creation operators in radial quantization mode. Then we can write the monomials in radial quantization as 
\begin{equation}
  \partial^{\mathbf{k}}\phi(0)|\Omega\rangle = \mathcal{N}_{\mathbf{k}}a_{\mathbf{k}}^\dagger|\Omega\rangle
\end{equation}
where
\begin{equation}
  a_{\mathbf{k}}^\dagger = a_{k_1}^\dagger a_{k_2}^\dagger... a_{k_n}^\dagger\;\;\;\mathcal{N}_{\mathbf{k}} = \bigg(\frac{1}{\sqrt{4\pi}}\bigg)^n \Gamma(k_1)...\Gamma(k_n)\sqrt{k_1...k_n}
\end{equation}

Now we can easily see how this speeds up the computation. Note that the in state and out states are completely orthogonal, i.e.
\begin{equation}
  \langle a_{\mathbf{k}}a_{\mathbf{k}'}^\dagger\rangle = ||\mathbf{k}||^2\delta_{\mathbf{k},\mathbf{k}'}
\end{equation}
where 
\begin{equation}
  ||\mathbf{k}||^2\equiv\prod_{k\in\mathbf{k}}\verb|BC|_k!=\frac{n!}{\operatorname{number\,of\,permutations\,of }\mathbf{k}},\;\;\; \delta_{\mathbf{k},\mathbf{k}'}\equiv \delta_{k_1k'_1}...\delta_{k_nk'_n} .
\end{equation}
Here $\verb|BC|_k$ is the bin count of $k$, or the number of times that $k$ show up in $\mathbf{k}$. Then we can obtain the Zamolodchikov metric $\mathfrak{g}_{\mathcal{OO'}}$ in a simple form 
\begin{equation}
  \mathfrak{g}_{\mathcal{OO'}} \equiv \langle \mathcal{O}(\infty)\mathcal{O}'(0)\rangle = \sum_{\mathbf{k}}\mathcal{C}_{\mathbf{k}}^{\mathcal{O}}\mathcal{C}_{\mathbf{k}}^{\mathcal{O}'}\mathcal{N}_{\mathbf{k}}^2||\mathbf{k}||^2
\end{equation}
Notice that in this case, we only have a single sum over $\mathbf{k}$ due to orthogonality, instead of a double sum over $\mathbf{k}$ and $\mathbf{k'}$ as in \eqref{eqn:wickexp}. 

To compute the matrix element of $\phi$ and $\phi^3$, we need to go through the following procedure. As an example, we would like to compute the $\phi^3$ matrix element
\begin{equation}
  \langle\mathcal{O}(x)\phi^3(y)\mathcal{O'}(z)\rangle
\end{equation}
The first step is to make a substitution
\begin{equation}
  \phi^3(y)\rightarrow\partial\phi(y_1)\partial\phi(y_2)\partial\phi(y_3)
\end{equation}
We made this substitution because $\partial\phi$ is a primary operator and a monomial, while $\phi$ is not. With the radial quantization decomposition, it is easy to compute a correlator of monomials, as we can see in the later steps; the fact that $\partial\phi$ is primary also allows us to perform a conformal transformation on the entire correlation function, which facilitates the computation. We can simply retrieve the original three-point function of $\phi^3$ by integrating over $y_i$:
\begin{equation}
  \langle\mathcal{O}(x)\phi^3(y)\mathcal{O'}(z)\rangle = \int dy_1 dy_2 dy_3 \langle\mathcal{O}(x)\partial\phi(y_1)\partial\phi(y_2)\partial\phi(y_3)\mathcal{O'}(z)\rangle\bigg|_{y=y_1=y_2=y_3}
  \label{ffphi3}
\end{equation}

Then we can decompose the right of \eqref{ffphi3} into monomials
\begin{equation}
  \langle \mathcal{O}(x)\partial\phi(y_1)\partial\phi(y_2)\partial\phi(y_3)\mathcal{O}'(z)\rangle = \sum_{\mathbf{k}\mathbf{k'}}\mathcal{C}_{\mathbf{k}}^{\mathcal{O}}\mathcal{C}_{\mathbf{k'}}^{\mathcal{O}'} G_{\mathbf{k}\mathbf{k'}}^{(\partial\phi\partial\phi\partial\phi)}(x,y_1,y_2,y_3,z)
\end{equation}
where 
\begin{equation}
  G_{\mathbf{k}\mathbf{k'}}^{(\partial\phi\partial\phi\partial\phi)}(x,y_1,y_2,y_3,z) \equiv \langle \partial^{\mathbf{k}}\phi(x)\partial\phi(y_1)\partial\phi(y_2)\partial\phi(y_3)\partial^{\mathbf{k'}}\phi(z)\rangle
\end{equation}
We then need to map it to the radial quantization matrix element through conformal transformation. Define
\begin{equation}
  G_{\mathbf{k}\mathbf{k'}}^{(\partial\phi\partial\phi\partial\phi)}(y_1,y_2,y_3) \equiv \langle \partial^{\mathbf{k}}\phi(\infty)\partial\phi(y_1)\partial\phi(y_2)\partial\phi(y_3)\partial^{\mathbf{k'}}\phi(0)\rangle
\end{equation}
then we have
\begin{multline}
  G_{\mathbf{k}\mathbf{k'}}^{(\partial\phi\partial\phi\partial\phi)}(x,y_1,y_2,y_3,z) \\\cong G_{\mathbf{k}\mathbf{k'}}^{(\partial\phi\partial\phi\partial\phi)}\left(\frac{y_1-z}{x-y_1},\frac{y_2-z}{x-y_2},\frac{y_2-z}{x-y_2}\right)\frac{(x-z)^{2-\Delta-\Delta'}}{(x-y_1)^2(x-y_2)^2(x-y_3)^2}
\end{multline}
where $\Delta$ is the dimension of an operator. We use the symbol $\cong$ because this equation is only true when the linear combination of $\mathbf{k}$ and $\mathbf{k'}$ add up to primary operators, since individual monomials does not have to obey conformal symmetry. Now we know that we can obtain $G_{\mathbf{k}\mathbf{k'}}^{(\partial\phi\partial\phi\partial\phi)}(x,y_1,y_2,y_3,z)$ by conformal transforming the radial quantization matrix elements $G_{\mathbf{k}\mathbf{k'}}^{(\partial\phi\partial\phi\partial\phi)}(y_1,y_2,y_3)$, we can compute these matrix elements by decomposing the monomials into creation and annihilation operators 
\begin{equation}
  G_{\mathbf{k}\mathbf{k'}}^{(\partial\phi\partial\phi\partial\phi)}(y_1,y_2,y_3) = \mathcal{N}_{\mathbf{k}}\mathcal{N}_{\mathbf{k'}}\langle a_{\mathbf{k}}\partial\phi(y_1)\partial\phi(y_2)\partial\phi(y_3)a^\dagger_{\mathbf{k'}}\rangle
\end{equation}
Then we need to decide whether to pull out a creation or an annihilation operator from $\partial\phi$. For $\phi^3$, there are 4 cases: $n\to n+3$, $n\to n+1$, $n\to n-1$, and $n\to n-3$. If we consider the $n\to n+1$ case for example, it means that there is one more $a^\dagger$ in $a^\dagger_{\mathbf{k'}}$ than $a$ in $a_{\mathbf{k}}$. To make the entire function non-vanishing when acting on the vacuum, we need two $a$ and one $a^\dagger$ from $(\partial\phi)^3$ to balance the left and the right. Then we can write
\begin{equation}
  G_{\mathbf{k}\mathbf{k'}}^{(\partial\phi\partial\phi\partial\phi)}(y_1,y_2,y_3) = \frac{\mathcal{N}_{\mathbf{k}}\mathcal{N}_{\mathbf{k'}}}{4\pi}\sum_{l_1,l_2,l_3=1}^\infty \sqrt{l_1l_2l_3}\,y_1^{-l_1-1}y_2^{-l_2-1}y_3^{l_3-1}\langle a_{\mathbf{k}}a_{l_1}a_{l_2}a^\dagger_{l_3}a^\dagger_{\mathbf{k'}}\rangle
\end{equation}
Then we can move $a_{l_3}^\dagger$ to the left and $a_{l_1}$, $a_{l_2}$ to the right, and contract with all possible $a_{k}$ and $a_{k'}^\dagger$. Note that we can ignore the delta function brought by $[a_l,a_l^\dagger]$ as it is removed by normal ordering of $\phi^3$ by definition. After these contractions we are left with 
\begin{equation}
  \frac{\mathcal{N}_{\mathbf{k}}\mathcal{N}_{\mathbf{k'}}}{4\pi}\sum_{k_i,k'_j,k'_l} \sqrt{k_ik'_jk'_l}\,y_1^{-k'_j-1}y_2^{-k'_l-1}y_3^{k_i-1}\langle a_{\mathbf{k}/k_i}a^\dagger_{\mathbf{k'}/k'_j,k'_l}\rangle
\end{equation}
Notice that the resulting expression is only non-zero when $\mathbf{k}/k_i=\mathbf{k'}/k'_j,k'_l$, which means that $\mathbf{k}$ and $\mathbf{k'}$ can only defer by 2 elements. One can show that $G_{\mathbf{k}\mathbf{k'}}^{(\partial\phi\partial\phi\partial\phi)}(y_1,y_2,y_3)$ is indeed a sum over terms of the form of 
\begin{equation}
  G_{k_ik_jk_l}^{(\partial\phi\partial\phi\partial\phi)}(y_1,y_2,y_3)\equiv \sqrt{k_ik'_jk'_l}\,y_1^{-k'_j-1}y_2^{-k'_l-1}y_3^{k_i-1}
\end{equation}
with $k_i-k'_j-k'_l = |\mathbf{k}|-|\mathbf{k'}| = \Delta-\Delta'$. If we want to transform the radial quantization matrix element back to the five-point function with finite position, we can perform the conformal transformation on each individual term 
\begin{multline}
  G_{k_ik_jk_l}^{(\partial\phi\partial\phi\partial\phi)}(x,y_1,y_2,y_3,z)\\\equiv \sqrt{k_ik'_jk'_l}\left(\frac{y_1-z}{x-y_1}\right)^{-k'_j-1}\left(\frac{y_2-z}{x-y_2}\right)^{-k'_l-1}\left(\frac{y_3-z}{x-y_3}\right)^{k_i-1}\frac{(x-z)^{2-\Delta-\Delta'}}{(x-y_1)^2(x-y_2)^2(x-y_3)^2}
\end{multline}
Now we have the expression for the five-point function with $(\partial\phi)^3$, and the next step will be to integrate over $y_i$ to obtain the matrix elements for $\phi^3$. The integration evaluates as 
\begin{equation}
  G_{k_ik'_jk'_l}^{(\phi^3)}(x,y,z) = \frac{\left(\left(\frac{x-y}{z-y}\right)^{k_i}-1\right)\left(\left(\frac{x-y}{z-y}\right)^{-k'_j}-1\right)\left(\left(\frac{x-y}{z-y}\right)^{-k'_l}-1\right)}{(-1)^{k_i-k'_j-k'_l}(x-z)^{\Delta+\Delta'}\sqrt{k_ik'_jk'_l}}
\end{equation}
when we take $y=y_1=y_2=y_3$. Now if we sum over possible $k_i$, $k'_j$, and $k'_l$, we can retrieve the $\phi^3$ matrix elements of monomials and therefore reconstruct the matrix element in primary basis. 

The next thing is to Fourier transform these elements to momentum space. For convenience, define 
\begin{equation}
  g_k(v) \equiv \frac{v^k-1}{\sqrt{k}}
\end{equation}
then the Fourier transform of $G_{k_ik'_jk'_l}^{(\phi^3)}(x,y,z)$ reads 
\begin{multline}
  \int_{-\infty}^{\infty} dxdydz\,e^{i(px-p'z)}(-1)^{k_i-k'_j-k'_l}(x-z-i\epsilon)^{-\Delta-\Delta'} \\\times g_{k_i}\left(\frac{x-y-i\epsilon}{z-y+i\epsilon}\right)g_{k'_j}\left(\frac{z-y+i\epsilon}{x-y-i\epsilon}\right)g_{k'_l}\left(\frac{z-y+i\epsilon}{x-y-i\epsilon}\right)
  \label{eq:RQFourier}
\end{multline}

To evaluate integrals of the form (\ref{eq:RQFourier}), first let us write them as
\begin{equation}
  \int_{-\infty}^{\infty} dxdydz\,e^{i(px-p'z)}(x-z-i\epsilon)^{-\Delta-\Delta'} F\left(\frac{x-y-i \epsilon}{z-y+i \epsilon} \right) .
\end{equation}
Making the following change of variables
\begin{equation}
  z=z'w+y,\;\;\;x=z'(w-1)+y,
\end{equation} 
 the integration becomes 
\begin{equation}
  2\pi\delta(p-p')\int dwdz'\,e^{iz'(qw-p)}(-z-i\epsilon)^{-\Delta-\Delta'}F\left(\frac{w-1-i\epsilon}{w+i\epsilon}\right),
\end{equation}
with $q\equiv p-p'$. The $z'$ integral evaluates to 
\begin{equation}
  4\pi\delta(p-p')\frac{2\pi^2}{\Gamma(\Delta+\Delta'-1)}\int dw\,(qw-p)^{\Delta+\Delta'-2}F\left(\frac{w-1-i\epsilon}{w+i\epsilon}\right) .
  \label{eqn:ffw0}
\end{equation}
Notice that $F\left(\frac{w-1-i\epsilon}{w+i\epsilon}\right)$ gives a pole at $w=-i\epsilon$ and $w=1+i\epsilon$. When we were integrating over $z'$, there was a term $e^{iz'qw}$. The $z'$ integral is performed by wrapping the branch cut from $z'=0$ to $z'=\infty$, therefore when we evaluate $e^{iz'qw}$, we always take $z'$ positive. An important assumption we made is that $p'>p$, which means that $q<0$. Therefore, to ensure that $e^{iz'qw}$ converges, we have to wrap the $w$ integral in the lower half plane, where $w$ goes to $-i\infty$. This translates to picking up the pole at $w=-i\epsilon$ when performing the $w$ integral. To evaluate the integral, we simply need to expand \eqref{eqn:ffw0} and extract the residue at $w=0$ (in the limit where $\epsilon\to0$).\footnote{We can also double-check that this prescription is the correct one by first expanding the factors of $g_k$ in (\ref{eq:RQFourier}) as a sum over a finite number of terms that vanish at $y\rightarrow \infty$ by expanding $v^k-1 = \sum_{n=1}^k\binom{k}{n}(v-1)^n $ and then doing the Fourier transforms term-by-term using (\ref{eq:FTAnalytic}).  } 

In our code, we used one more step to speed up the computation. Notice that $F\left(v\right)$ is indeed a product of $g_k\left(v\right)$, which can be expanded as a sum in the form of
\begin{equation}
  g_a(v)g_b(v)\sim(v^a-1)(v^b-1)=(v^{a+b}-1)-(v^a-1)-(v^b-1).
\end{equation}
Now instead of expanding a product of powers of $v$, we only need to consider the Fourier transform of a single power of $v$ each time, which reduces the processing time of residue computation.

In summary,  we now have the following complete recipe to compute matrix elements of $\phi^3$: First, we evaluate $G_{k_ik_jk_l}^{(\partial\phi\partial\phi\partial\phi)}(y_1,y_2,y_3)$ for possible contractions $k_i$, $k'_j$, and $k'_l$ between two monomials, then transform it to the five-point function in position space $G_{k_ik_jk_l}^{(\partial\phi\partial\phi\partial\phi)}(x, y_1,y_2,y_3,z)$. Then we integrate over $y_i$ to obtain the $\phi^3$ elements from $(\partial\phi)^3$, and Fourier transform the position space correlator to momentum space. Lastly, we sum over possible contractions to get $\langle\partial^{\mathbf{k}}\phi(p)\phi^3(0)\partial^{\mathbf{k'}}\phi(p')\rangle$, and then sum over the monomials with appropriate coefficients to get the matrix elements of $\phi^3$ in primary basis.

\section{Cross-checking with Feynman diagram results}
\label{sec:Feynman}

In this section, we will present the truncation results in radial quantization, compared with first order and second order corrections from Feynman diagrams assumed weak coupling. We will first compare the functions of form factor in the regime where $s<4$. In this regime, the function is smooth, so that we can avoid the comparison between poles shifted by truncation effects, and observe a better agreement between truncation results and Feynman diagram results. By analytic continuation, if the results from truncation agrees with the analytic solution in the negative $s$ regime, they should also agree in the positive $s$ regime. For the one-loop result, we will also compare the integrated results in the positive $s$ regime. From now on, we will take units of $m_0=1$ to avoid clutter.

By old-fashioned time-independent perturbation theory, we can expand the form factor to leading orders
\begin{equation}
  \frac{\langle \Omega | T_{--} |\mu\rangle\langle\mu| A |p_1\rangle}{s-\mu^2}\approx\frac{\langle \Omega | T_{--} |\mu^{(0)}+\lambda\mu^{(1)}\rangle\langle\mu^{(0)}+\lambda\mu^{(1)}|\left(\delta m^2\phi + \frac{\lambda}{6}\phi^3\right)|p_1^{(0)}+\lambda p_1^{(1)}\rangle}{s-(\mu^{(0)2}+\lambda\mu^{(1)2})},
  \label{eq:pertexp}
\end{equation}
where $\mu^{(0)}$ is the unperturbed eigenstate of the Hamiltonian, and $\mu^{(1)}$ is the first order correction, and the same notation applies to $p_1$. The definition of $A$ is $A(x) \equiv \delta m^2 \phi + \frac{\lambda}{6} \phi^3$. 
 Given that $\delta m^2 = \frac{3}{2}\frac{\lambda^2}{(4!)^2} + \mathcal{O}(\lambda^3)$, we can extract the first order and second order corrections to the form factor, as we will present in more detail in the subsections. 

When we want to compare the form factor with negative $s$, there are two ways to do it: the first way is to read off the residue of the pole from one-particle eigenstates, as presented in Equation \eqref{eq:TChannel}, which we will call the $t$-channel method. Another way is to use the dispersion relation
\begin{equation}
  \Delta(s)=-\frac{1}{\pi}\int_{4m^2}^{\infty} d\mu^2 \frac{\operatorname{Im}(\Delta(\mu^2))}{s-\mu^2+i\epsilon}
  \label{eq:dispersion}
\end{equation}
to reconstruct the entire form factor from just the imaginary piece in the positive $s$ regime. Taking the imaginary part of form factor gives a delta function 
\begin{equation}
  \operatorname{Im}\,\frac{\langle \Omega | T_{--} |\mu\rangle\langle\mu|A(p_2)|p_1\rangle}{s-\mu^2+i\epsilon} = -\pi \langle \Omega | T_{--} |\mu\rangle\langle\mu|A(p_2)|p_1\rangle\delta(s-\mu^2),
  \label{eq:Fdeltapiece}
\end{equation}
and in truncation, taking the integral on the right side of Equation \eqref{eq:dispersion} gives an expression for $\tilde{F}(s)$ that can be evaluated for any complex value of $s$:
\begin{equation}
  \tilde{F}(s)= 
\sum_i \frac{\langle \Omega | T_{--} |\mu_i\rangle\langle\mu_i|A(p_2)|p_1\rangle_{(p_1+p_2)^2 = \mu_i^2}}{s-\mu_i^2+i\epsilon}
  \label{eq:FImaginary}
\end{equation}
where $\mu_i^2$ are the eigenvalues of the Hamiltonian. By contrast, our original expression (\ref{eq:FtildeFinal}), which took $(p_1+p_2)^2=s$, was restricted to the physical regime $s>4m^2$. When we take the imaginary part, the $\delta$ function in (\ref{eq:Fdeltapiece}) enforces the on-shell condition $(p_1+p_2)^2=\mu_i^2$ and allows us to fix the value of $p_1$. For this reason, when we use equation  (\ref{eq:FImaginary}) to compute the form factor at negative $s$, we have to also choose $p_1$ to take different values for each term in the sum.

\subsection{One-loop check}

\begin{figure}[t!]
  \centering
  \includegraphics[width=0.5\textwidth]{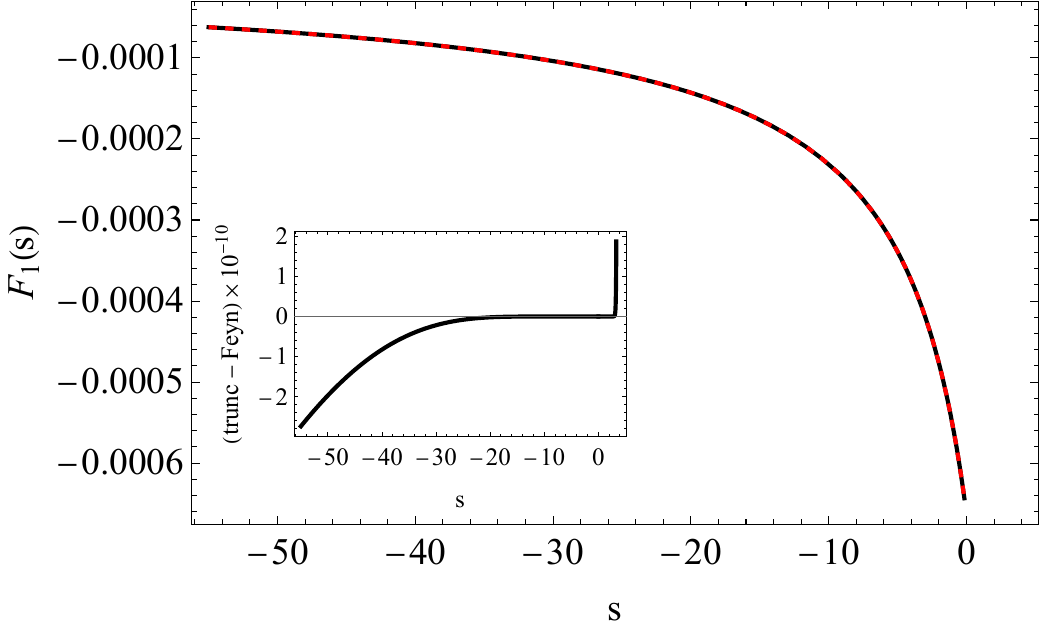}
  \caption{Plot of one-loop result from Feynman diagram (black solid line) and truncation (red dashed line, at $\Delta\operatorname{max}=20$) in negative $s$ regime. The residual plot shows the difference between the two lines.}
  \label{fig:1loopnegs}
\end{figure}

From Equation \eqref{eq:pertexp}, we can extract the first order correction to the form factor 
\begin{equation}
  \frac{1}{6}\frac{\langle \Omega | T_{--} |\mu^{(0)}\rangle\langle\mu^{(0)}|\phi^3|p_2^{(0)}\rangle}{s-\mu^{(0)2}} .
\end{equation}
\begin{figure}[t!]
  \centering
  \includegraphics[width=0.4\textwidth]{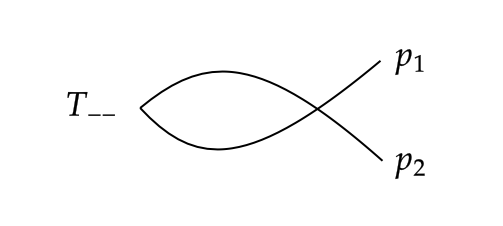}
  \includegraphics[width=0.59\textwidth]{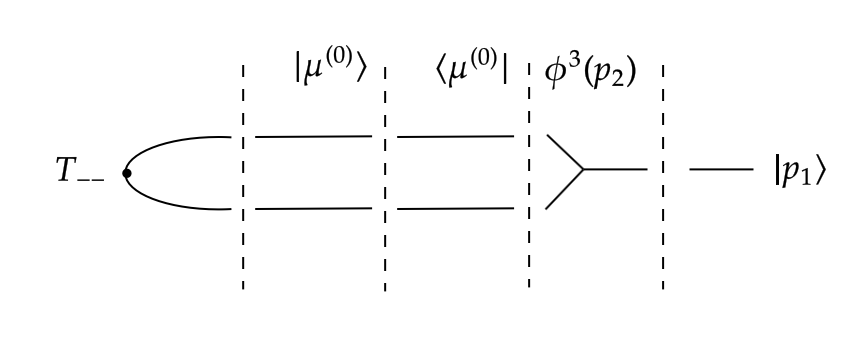}
  \caption{One loop Feynman diagram of $\langle T_{--}|p_1p_2\rangle$ (left) and first order correction in truncation $\langle\Omega|T_{--}|\mu\rangle\langle\mu|A(p_1)|p_2\rangle$ (right) .}
  \label{fig:1loopdiagram}
\end{figure}
In Fig.~\ref{fig:1loopdiagram}, we show the first order corrections with and without applying LSZ in truncation. Notice that in truncation, we only need to consider two-particle intermediate states for $\mu^{(0)}$, and only the $n\to n-1$ piece of $\phi^3$ interaction will contribute. The one-loop Feynman diagram gives the $\CO(\lambda)$ contribution $F_1$ to the form factor ($F \supset \lambda F_1$):
\begin{equation}
  F_1(s)= -\frac{1}{8\pi} \left(-1+\frac{4\tan^{-1}\left(\frac{\sqrt{s}}{\sqrt{4-s}}\right)}{\sqrt{s(4-s)}}\right)/s,
\end{equation}
and we can compare this function with the truncation result in the negative $s$ regime, as shown in Fig.~\ref{fig:1loopnegs}. We can see that the two lines are almost in exact agreement, with a difference as low as $10^{-10}$. 

Next, we would also want to check the agreement in positive $s$ regime with this method. In truncation, the eigenvalues of the Hamiltonian are always discrete, therefore it is impossible to reconstruct the poles perfectly, and the position of the poles are so sensitive to truncation effects that we cannot compare them with the Feynman diagram result. Rather than looking at a function with poles in positive $s$, we can compare the integrated imaginary part of the form factor:
\begin{equation}
  \int_0^{s} ds' \operatorname{Im}F_1(s') 
  = -\pi \sum_{\mu_n^2<s}\langle \Omega | T_{--} |\bar{\mu}_n\rangle\langle\bar{\mu}_n|A(p_2)|p_1\rangle
\end{equation}
where $\bar{\mu}_n$ means that we require the intermediate state to be on-shell.
 In truncation, we now have a sum over step functions which are less sensitive to truncation effects.

\begin{figure}[t!]
  \centering
  \includegraphics[width=0.5\textwidth]{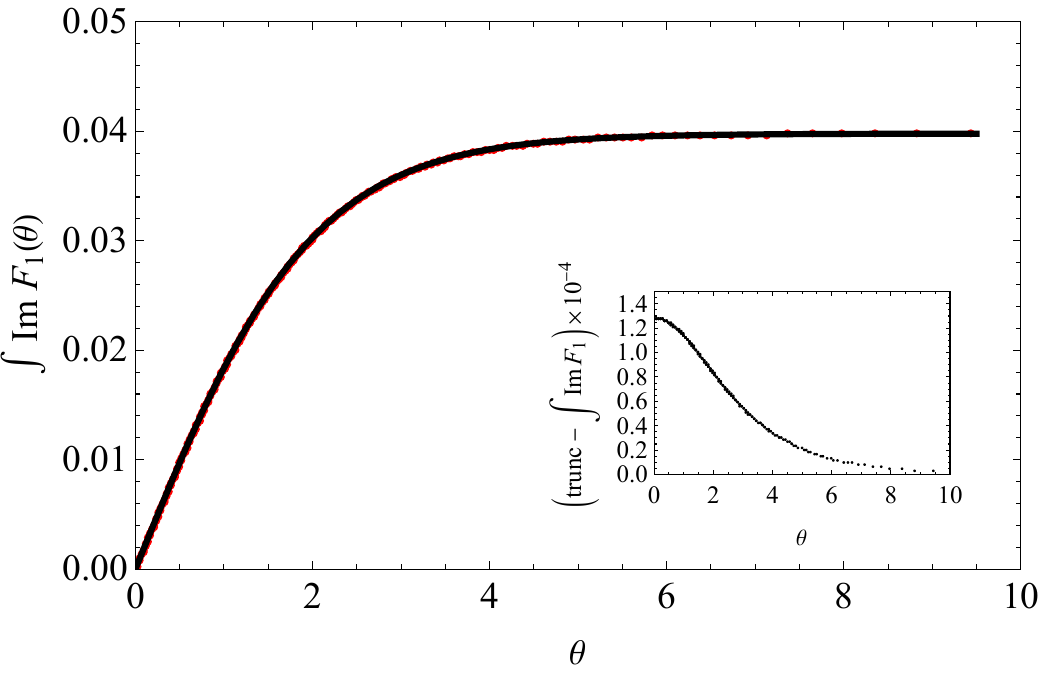}
  \caption{Plot of $\int_0^s ds' F_1(s')$ comparing the analytic solution (black solid line) and truncation result from Jacobi method (red dashed line, included 484 intermediate states) in positive $s$ regime. The residual plot shows the difference between the two lines.}
  \label{fig:1loopposs}
\end{figure}

Indeed, we can exploit the fact that we only need to consider two-particle inserted states, and generate an analytic result for the one-loop correction.  The closed-form expression for the two-particle eigenvalues in the free theory are \cite{Anand:2020gnn}
\begin{equation}
  \mu_j = 2\sec\left(\frac{(2j+1)\pi}{2\Delta_{\operatorname{max}}+1}\right) + \CO(\Delta_{\rm max}^{-1}) .
\end{equation}
Moreover, although we do not have an analytic derivation, it is easy to verify empirically that the following relation holds approximately at large $\Delta_{\rm max}$, up to higher order in $1/\Delta_{\rm max}$ corrections:
\begin{equation}
  \langle \Omega|T_{--}(0)|\mu_i\rangle\langle\mu_i|\frac{1}{6} \phi^3|p_1\rangle \approx \frac{1}{2(2\Delta_{\rm max}+1)} \frac{1}{\mu_i}. 
\end{equation}
At this order, the matrix elements $\< \mu_i | \phi^3(q) | p_1\>$ do not depend on $q$.  Using the large $\Delta_{\rm max}$ expression for the eigenvalues $\mu_j$, we can analytically do the sum over eigenstates to evaluate the form factor:
\begin{equation}
\tilde{F}_{T_{--}}(s) \approx  \lambda \sum_i \frac{1}{(2\Delta_{\rm max}+1)} \frac{1}{\mu_i} \frac{1}{s-\mu_i^2 + i \epsilon} \approx  \lambda \int_{4}^\infty d \mu^2 \frac{1}{2\pi \mu^3 \sqrt{\mu^2-4}} \frac{1}{s-\mu^2+i\epsilon},
\label{eq:OneLoopAnalytic}
\end{equation}
where we have used $d\mu/dj = \frac{\pi \mu \sqrt{\mu^2-4}}{2\Delta_{\rm max}+1}$. Taking the imaginary part of this expression, we see that it
matches the Feynman diagram result
\begin{equation}
  \operatorname{Im} F_1(s) =
-  \frac{1}{4s\sqrt{s(s-4)}}.
\end{equation}
A comparison between analytic and truncation results in the positive $s$ regime is shown in Fig.~\ref{fig:1loopposs}. With 484 intermediate states involved, we can see that the truncation is successfully reproducing the analytic result, which is another check that applying LSZ in Hamiltonian truncation is a valid method.

\subsection{Two-loop check}

\begin{figure}[t!]
  \centering
  \includegraphics[width=0.48\textwidth]{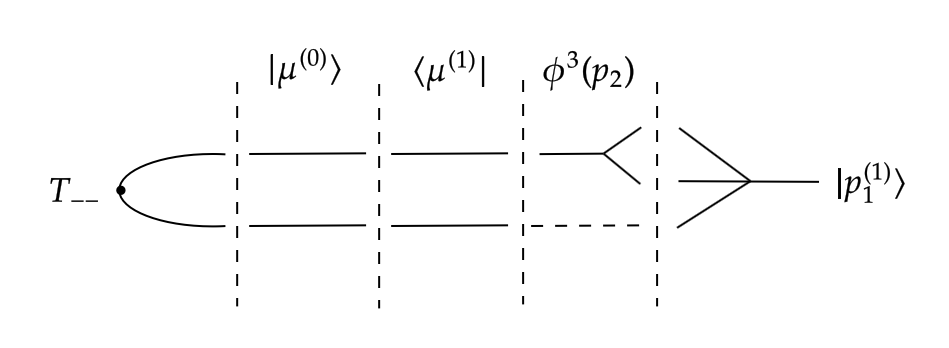}
  \includegraphics[width=0.48\textwidth]{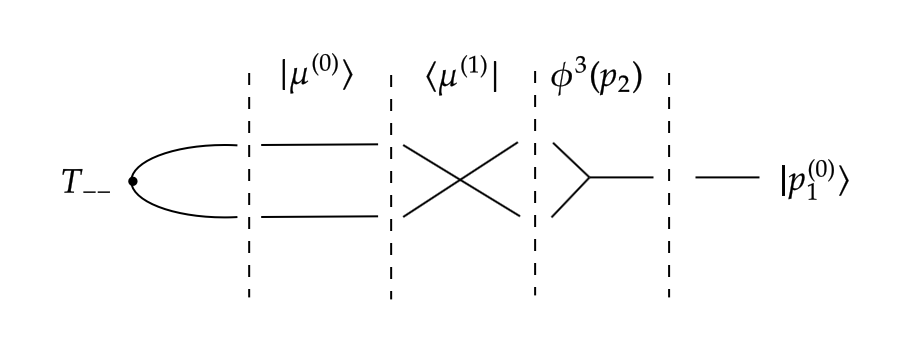}
  \includegraphics[width=0.48\textwidth]{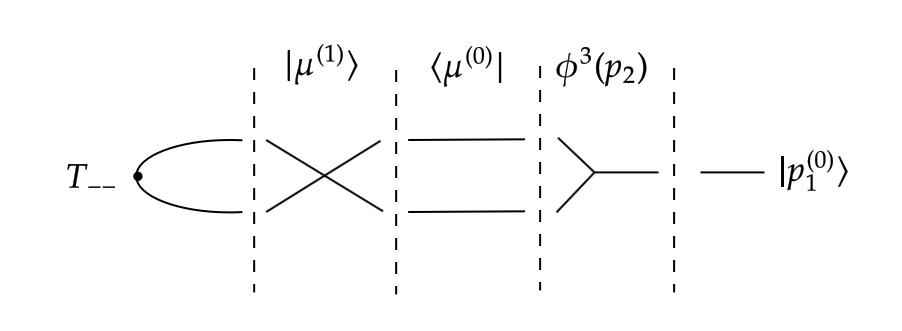}
  \includegraphics[width=0.48\textwidth]{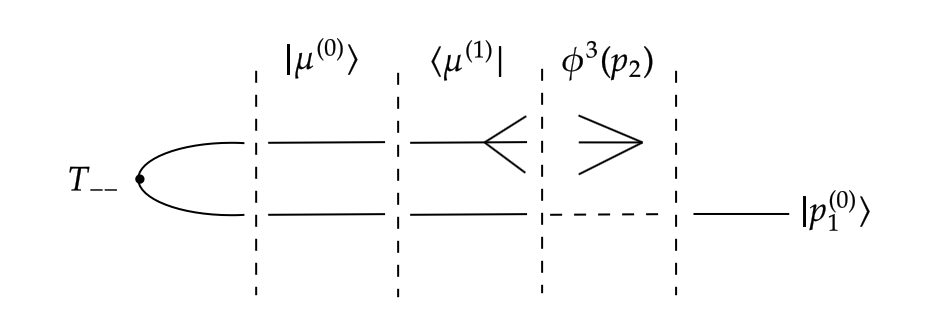}
  \includegraphics[width=0.48\textwidth]{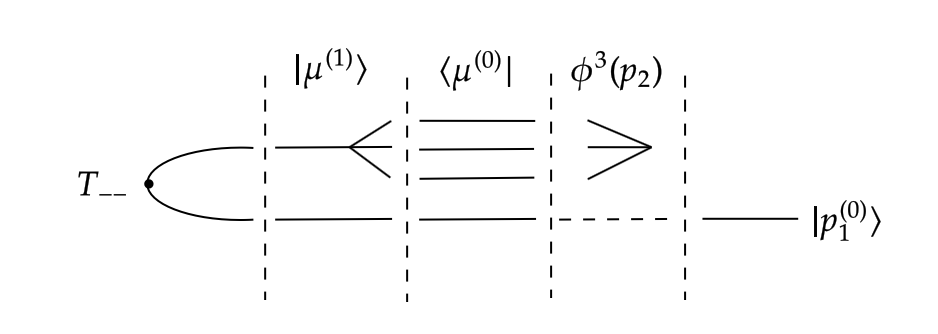}
  \includegraphics[width=0.48\textwidth]{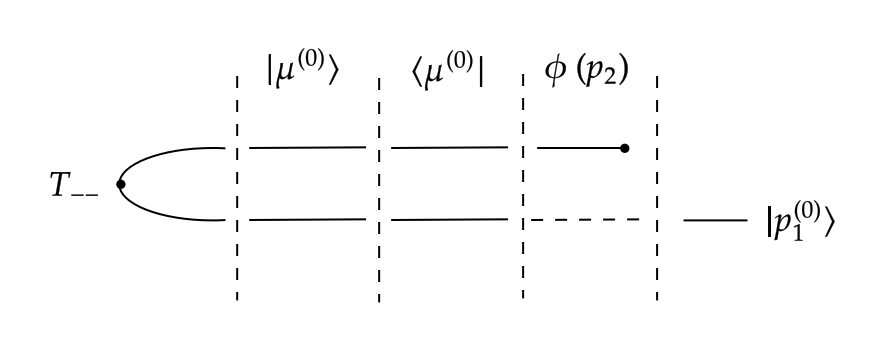}
  \caption{Second order corrections in truncation.}
  \label{fig:2looptrunc}
\end{figure}

From Equation \eqref{eq:pertexp}, the second order corrections to the numerator are 

\begin{equation}
\begin{aligned}
&  \frac{3}{2}\frac{\lambda^2}{(4!)^2}\langle \Omega | T_{--} | \mu^{(0)} \rangle \langle \mu^{(0)}|\phi|p_1^{(0)}\rangle,
  \label{eq:2ndorder1} \\
 & \frac{\lambda^2}{6}\langle \Omega | T_{--} | \mu^{(1)} \rangle \langle \mu^{(0)}|\phi^3|p_1^{(0)}\rangle, \\
&  \frac{\lambda^2}{6}\langle \Omega | T_{--} | \mu^{(0)} \rangle \langle \mu^{(1)}|\phi^3|p_1^{(0)}\rangle, \\
&   \frac{\lambda^2}{6}\langle \Omega | T_{--} | \mu^{(0)} \rangle \langle \mu^{(0)}|\phi^3|p_1^{(1)}\rangle, 
  \end{aligned}
\end{equation}
and their corresponding diagrams are shown in Fig.\ref{fig:2looptrunc}.

\begin{figure}[t!]
  \centering
  \includegraphics[width=\textwidth]{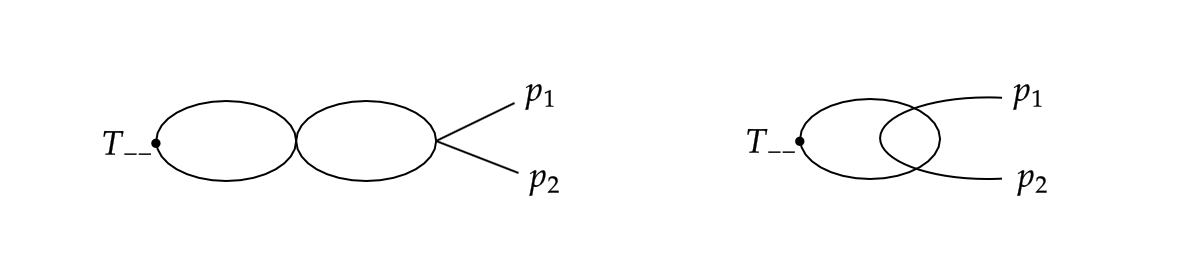}
  \caption{Two-loop Feynman diagrams contributing to $\<\Omega | T_{--} | p_1, p_2\>$.}
  \label{fig:2loopdiagram}
\end{figure}

In addition to the quantities in \eqref{eq:2ndorder1}, we also need to consider the correction from the denominator:
\begin{equation}
  \frac{1}{s-(\mu^{(0)2}+\lambda\mu^{(1)2})} \approx \frac{1}{s-\mu^{(0)2}}+\lambda\mu^{(1)2}\frac{1}{(s-\mu^{(0)2})^2}.
\end{equation}
The complete second order correction reads 
\begin{equation}
  \begin{aligned}
    &\frac{1}{6}\frac{\langle \Omega | T_{--} | \mu^{(1)} \rangle \langle \mu^{(0)}|\phi^3|p_1^{(0)}\rangle+\langle \Omega | T_{--} | \mu^{(0)} \rangle \langle \mu^{(1)}|\phi^3|p_1^{(0)}\rangle+\langle \Omega | T_{--} | \mu^{(0)} \rangle \langle \mu^{(0)}|\phi^3|p_1^{(1)}\rangle}{s-\mu^{(0)2}} \\
  &-(\delta m^2)^{(2)} \langle \Omega | T_{--} | \mu^{(0)} \rangle \langle \mu^{(0)}|\phi|p_1^{(0)}\rangle
  +\frac{\mu^{(1)2}}{6}\frac{\langle \Omega | T_{--} |\mu^{(0)}\rangle\langle\mu^{(0)}|\phi^3|p_2^{(0)}\rangle}{(s-\mu^{(0)2})^2}.
  \end{aligned}
  \label{eqn:2loopff}
\end{equation}

There are two new qualitative features about the form factor at second order as compared to first order.  The first is that now there are four-particle states that must be included in the sums.  This is not a significant conceptual difference, but it does make it more difficult to obtain analytic calculations of the form factor in truncation since the four-particle states are more complicated than the two-particle ones, and it also means the maximum $\Delta_{\rm max}$ we can take in practice will be lower since the number of four-particle states grows more rapidly with $\Delta_{\rm max}$.  

 More significantly, at second order there is no a contribution both from the $\phi$ and the $\phi^3$ term in the equation of motion.  This fact introduces an interesting subtlety.  Because we evaluate the form factors on shell, LSZ tells us that in the exact second order form factors, the contributions from $\phi$ and $\phi^3$ individually will be diverge due to the $1/(p^2-m^2)$ pole.  Instead, it is only the sum of these two terms as they appear in the equation of motion that will remain finite on-shell.  Since we are evaluating the form factor with a finite $\Delta_{\rm max}$, in our case the contributions from $\phi$ and $\phi^3$ will remain finite on-shell; indeed, this is our main purpose for using the equations of motion.  As $\Delta_{\rm max}$ is taken to $\infty$, we should see the $\phi$ and $\phi^3$ terms individual grow without bound.  It is important to actually check in practice whether or not this divergence, combined with other possible truncation effects, correctly cancels out in the large $\Delta_{\rm max}$ limit.  In fact, it is not enough for the divergence to merely cancel -- it is crucial that the subleading finite terms also produce the correct continuum limit result and are not spoiled by subtle residual truncation errors due to the leading divergence.\footnote{We thank Jed Thompson and Matthew Walters for emphasizing this issue to us.}

\begin{figure}[t!]
  \centering
  \includegraphics[width=0.8\textwidth]{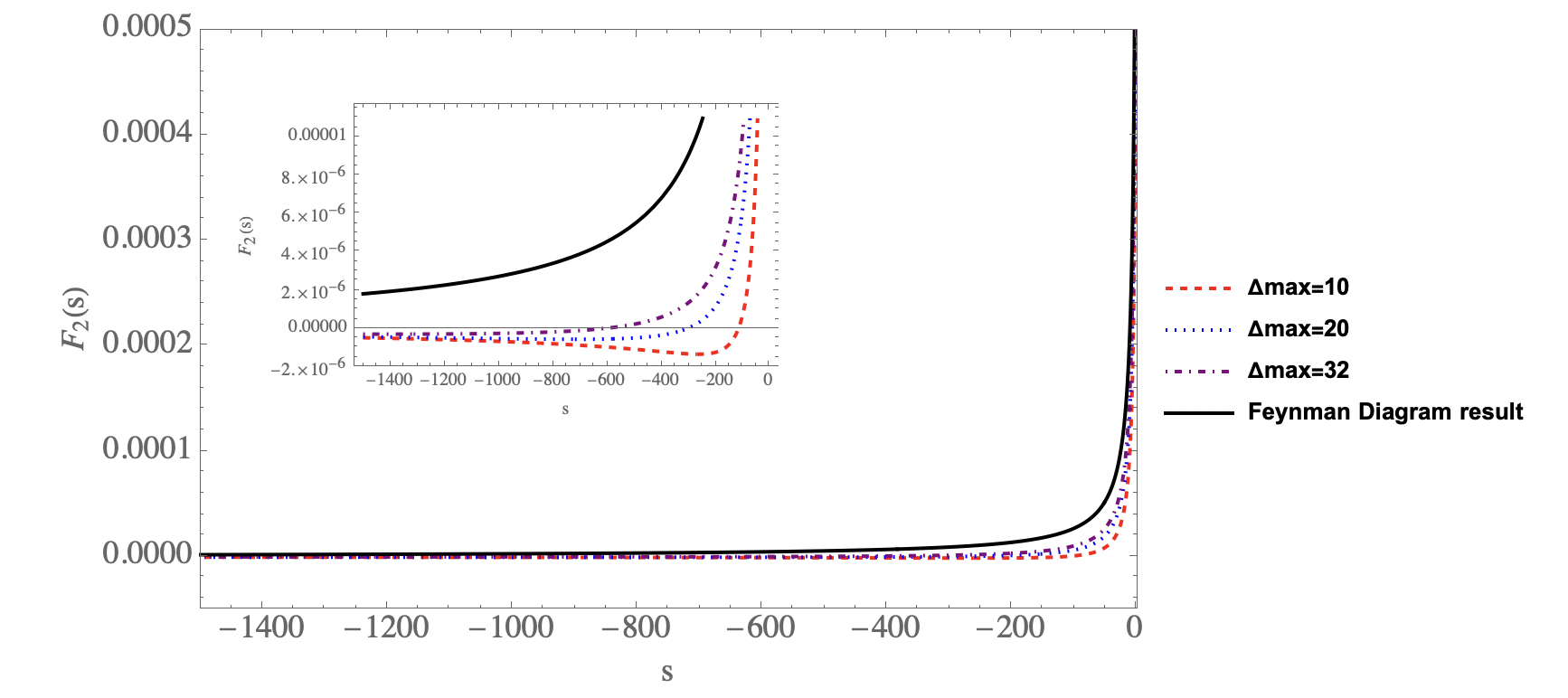}
  \caption{Comparison between truncation and Feynman diagram 2-loop results, in negative $s$ regime. The inset plot shows the same function at different scale, and one can see that for finite $\Delta_{\rm max}$, at sufficiently large $s$ the truncation approximation eventually breaks down.}
  \label{fig:2loopnegs}
\end{figure}

Plotting the imaginary part of the second order result at positive $s$ is challenging because in truncation, the imaginary part is a sum over $\delta$ functions and the second order piece therefore involves derivatives of $\delta$ functions. 
Instead, we will focus on the real part of the second order result in the negative $s$ regime. In Fig.\ref{fig:2loopnegs}, we present the comparison between truncation results after assembling the $\phi$ and $\phi^3$ pieces as presented in Equation \eqref{eqn:2loopff}. The agreement is not as good as the one-loop result, but we can still see the convergence towards the Feynman diagram result as we increase the truncation dimension. 

\begin{figure}[t!]
  \centering
  \includegraphics[width=0.8\textwidth]{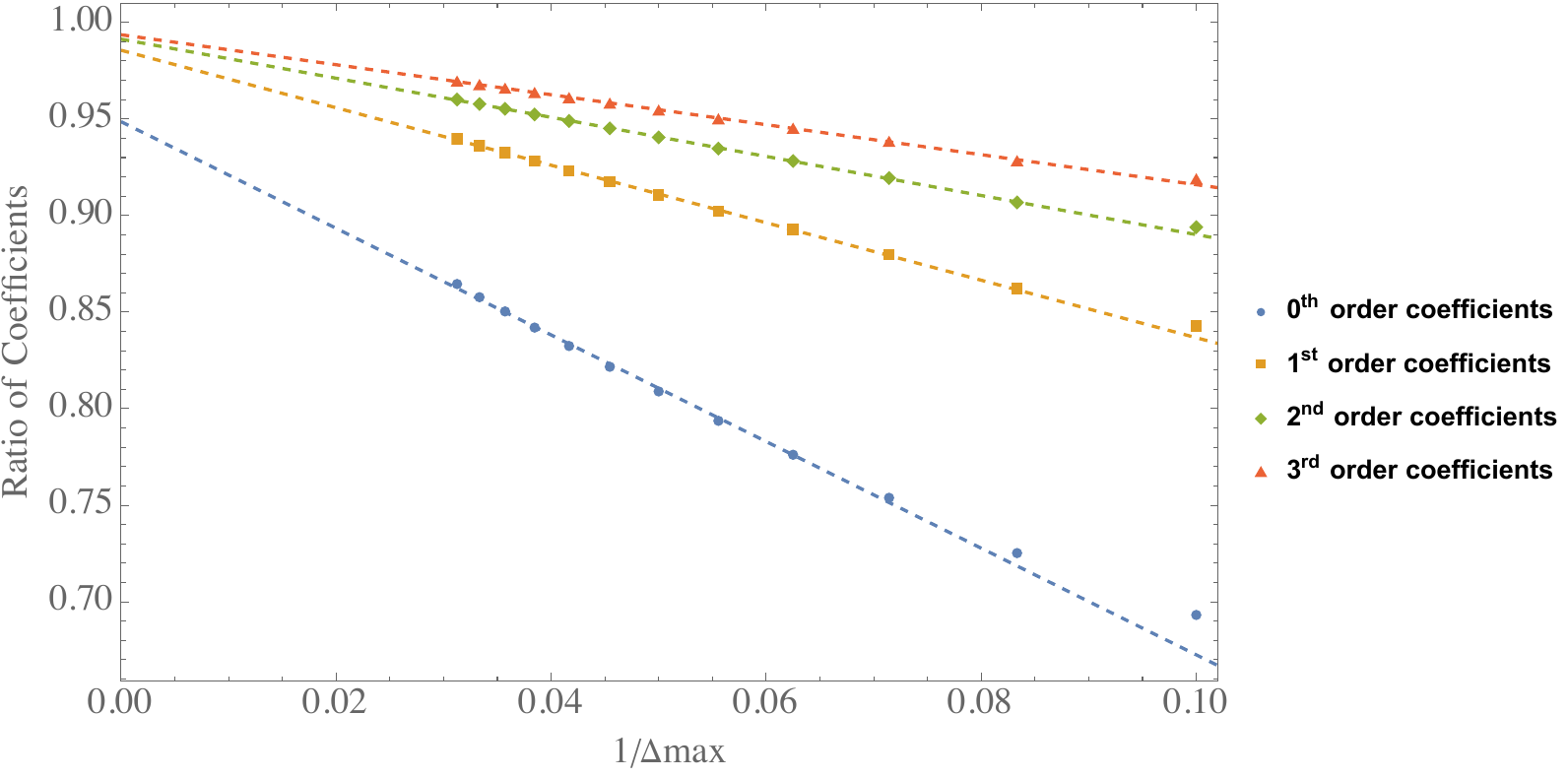}
  \caption{Ratio between Taylor series coefficients of truncation result and Feynman diagram results, expanded at $s=0$. The dashed lines are linear fits to Taylor series coefficients of different orders. }
  \label{fig:2loopffcoef}
\end{figure}

\begin{figure}[t!]
  \centering
  \includegraphics[width=0.49\textwidth]{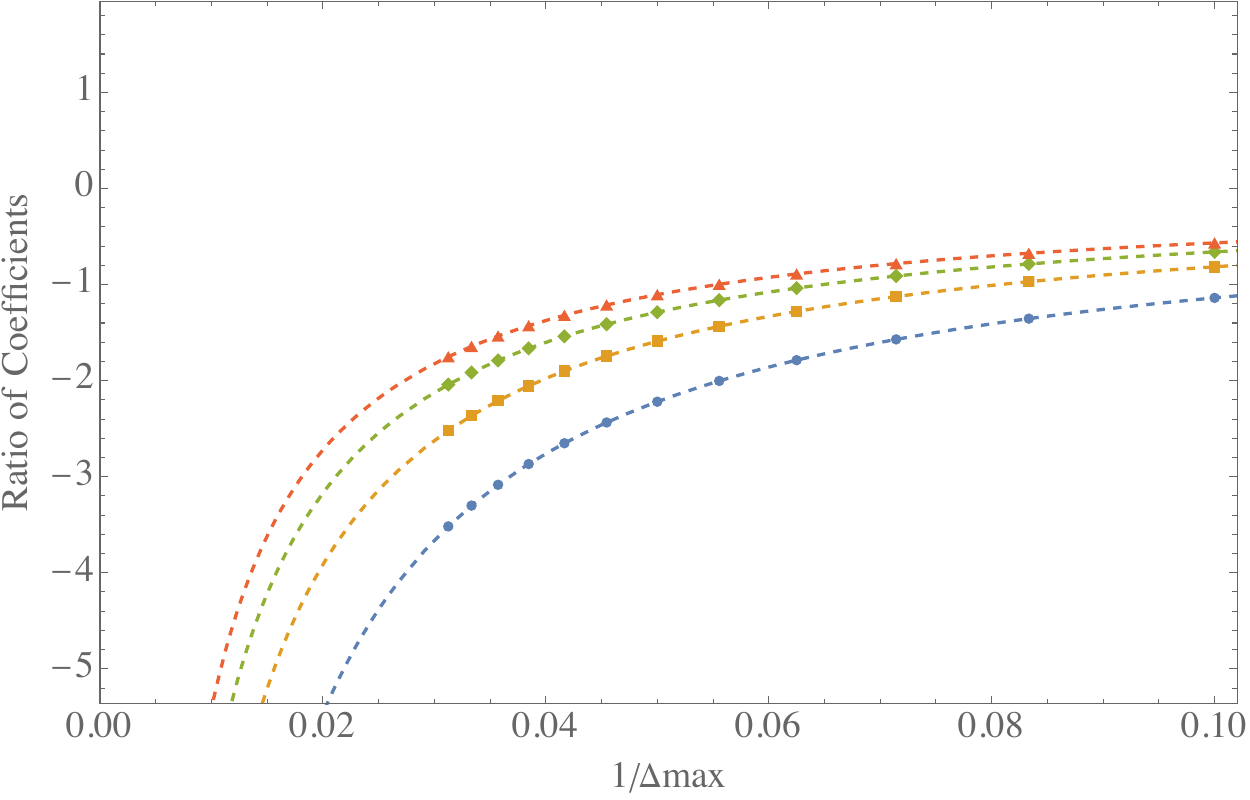}
  \includegraphics[width=0.49\textwidth]{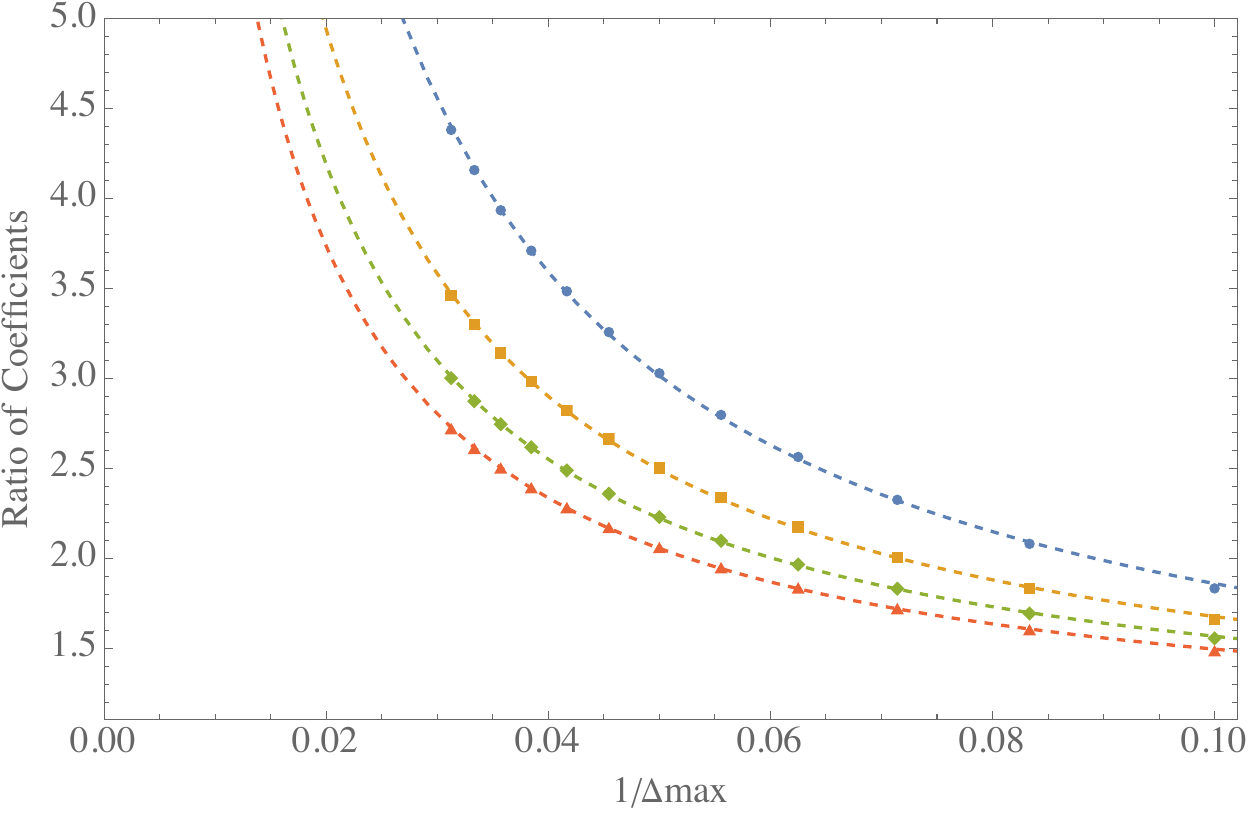}
  \caption{Quotient of Taylor series coefficients of $\phi$ (left) and $\phi^3$ (right) contributions with the 2-loop correction coefficients from Feynman diagram; color scheme for the Tayor series coefficients is the same as in Fig.~\ref{fig:2loopffcoef}. 
  The dashed lines fits $a \Delta_{\rm max}+b$ to Taylor series coefficients; the divergent trend as $1/\Delta_{\rm max} \rightarrow 0$ is clearly visible.}
  \label{fig:2loopcompcoef}
\end{figure}

In Fig.\ref{fig:2loopffcoef}, we present the quotient of the first four Taylor series coefficients of truncation and Feynman diagram results, and their quotient converges to 1 as $1/\Delta_{\rm max}$. In Fig.\ref{fig:2loopcompcoef}, we present the $\phi$ and $\phi^3$ contributions to the form factor, and both of them diverge as $\Delta_{\rm max}$. This is the aforementioned result of the cancellation between $\phi$ and $\phi^3$ pieces, as each term individually is growing while their sum is converging. 

In Fig.\ref{fig:2loopffcoef}, we can see that the higher order terms from truncation are closer to the analytic results. To explain this, when we Taylor expand the form factor, it can be written as 
\begin{equation}
  \sum_i \frac{f_i}{s-\mu_i^2} = \sum_{n=0}^\infty s^n  \left( \sum_i \frac{-f_i}{\mu_i^{2+2n}}\right) , 
 \;\;\;  f_i\equiv\langle T_{--}|\mu_i\rangle\langle\mu_i|\delta m^2\phi+\frac{\lambda}{6}\phi^3|p_1\rangle.
\end{equation}
Notice that the term constant in $s$ has the lowest power of $\mu_i$ in the denominator, therefore is the most sensitive to the truncation effect. In constrast, $\mu_i$ in the denominator has higher power in the higher order terms, therefore the truncation effects are suppressed. In the same truncation dimension, we always see that higher order Taylor coefficients are closer to the expected value.

Finally, let us look in more detail at the contribution from the $\phi(q)$ to the two-loop form factor. In the large truncation limit,\footnote{The fact that this relation holds in the large truncation limit is easy to derive by using the fact that the two-particle eigenstates simply approach $\delta$ functions in the momentum fraction of the individual particle momenta.  Surprisingly, we have found empirically that (\ref{eq:Tmmphi}) holds at any finite truncation in the free, massive theory,  though we have not been able to derive this analytically. }
  \begin{equation}
 \< T_{--} | \mu \> \< \mu | \phi(q) | p_1\> = p_1.
 \label{eq:Tmmphi}
 \end{equation}
 A simple consequence is that after combining the $p_1 \leftrightarrow p_2$ contributions, 
 \begin{equation}
 \< T_{--} | \mu \> \< \mu | \phi(q) | p_1\>+ \< T_{--} | \mu \> \< \mu | \phi(q) | p_2\> = p_1+p_2=1.
 \end{equation}
 Following a similar derivation to the one leading to (\ref{eq:OneLoopAnalytic}), the contribution to the form factor is
 \begin{equation}
 \tilde{F} = \sum_i \frac{\delta m^2}{s-\mu_i^2+i\epsilon}
 \approx \delta m^2\frac{(2\Delta_{\rm max}+1)}{2s} \left( \frac{1}{2} - \frac{\sqrt{4-s}}{4-s}\right)
 \end{equation}
so in this case, we can explicitly see that the contribution from $\phi(q)$ alone diverges like $\CO(\Delta_{\rm max})$ at large $\Delta_{\rm max}$.

\section{Strong Coupling Result}
\label{sec:strong}

\begin{figure}[t!]
  \centering
  \includegraphics[width=0.4\textwidth]{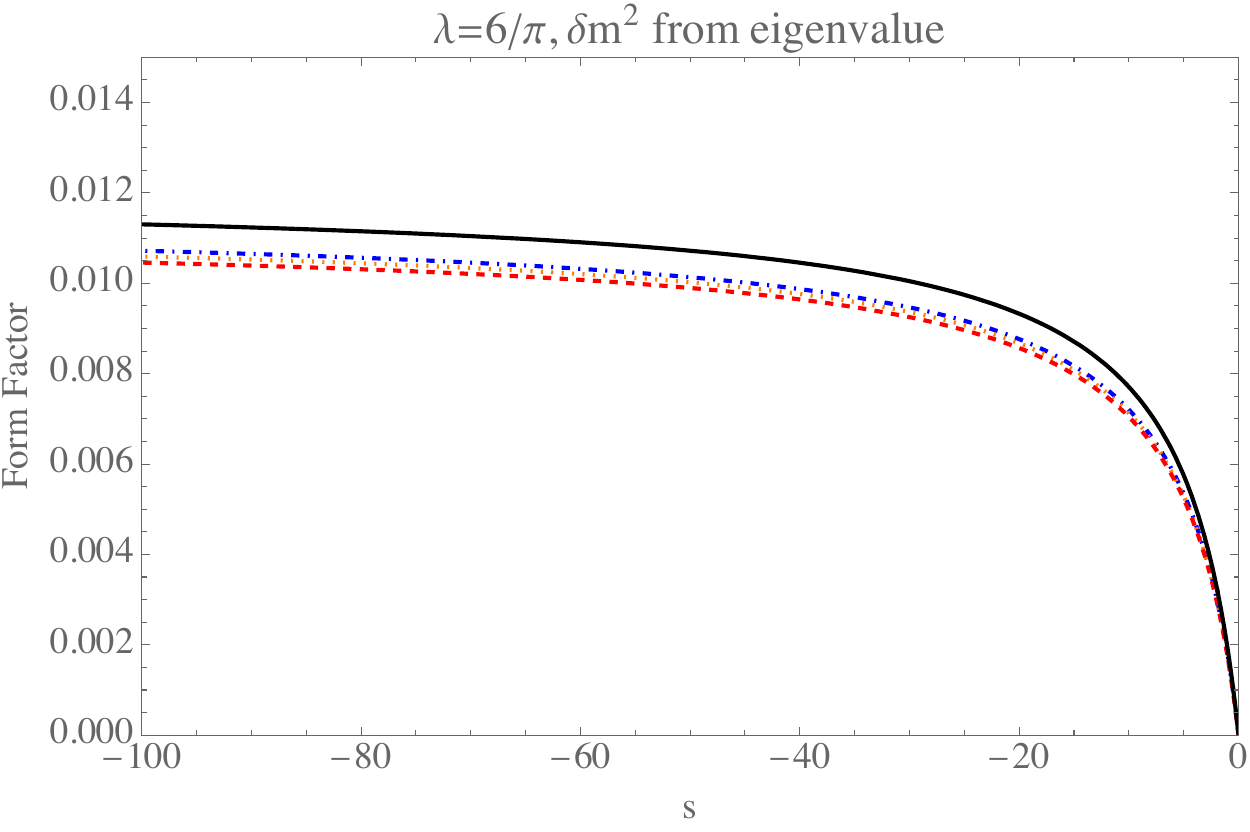}
  \includegraphics[width=0.49\textwidth]{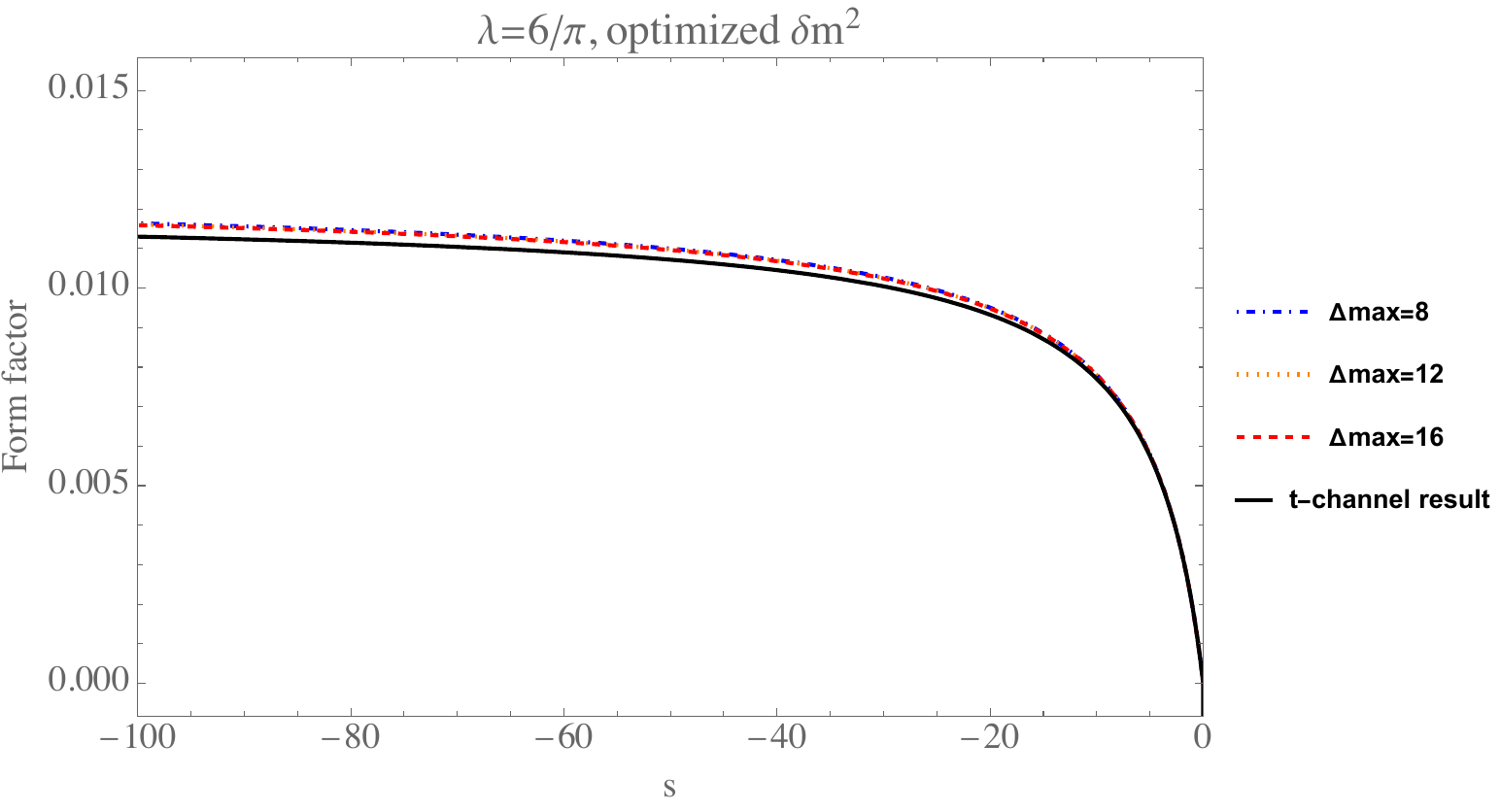}
  \includegraphics[width=0.4\textwidth]{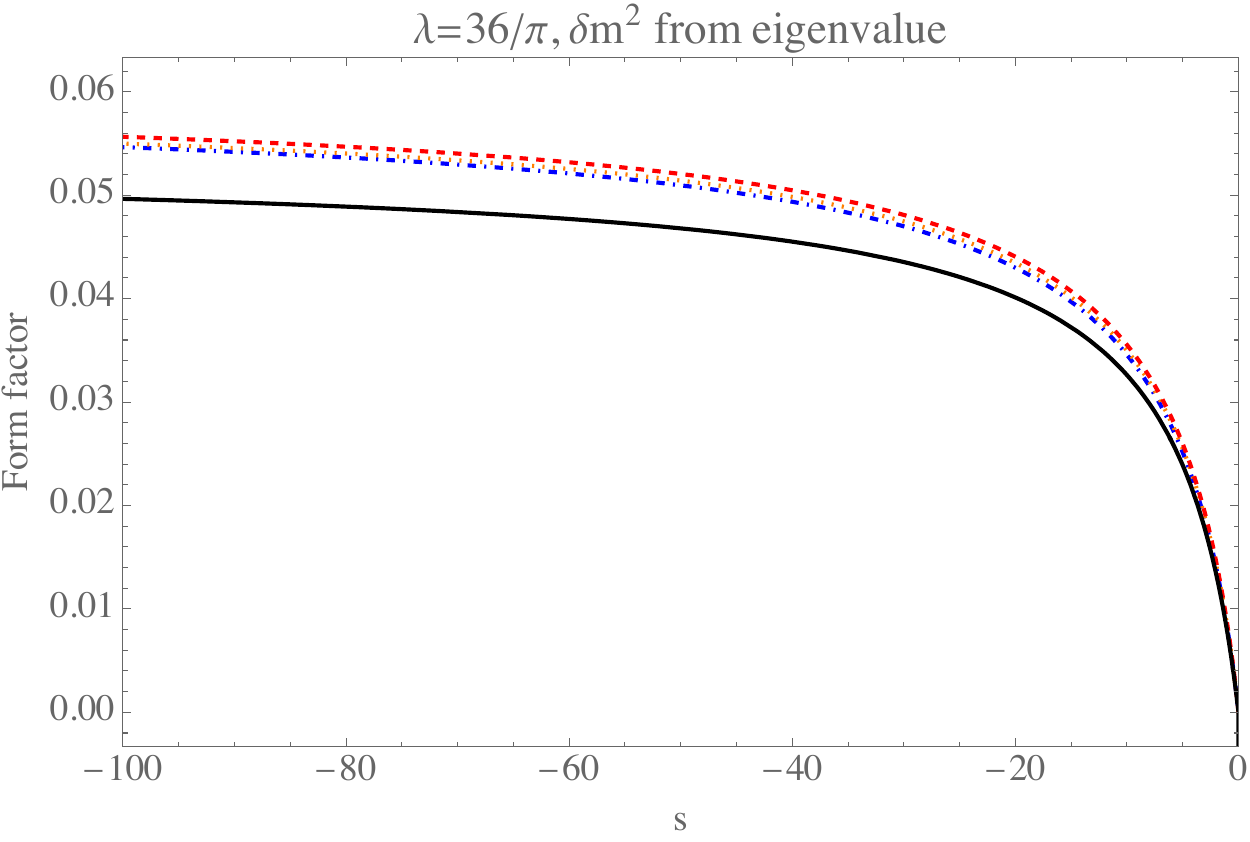}
  \includegraphics[width=0.49\textwidth]{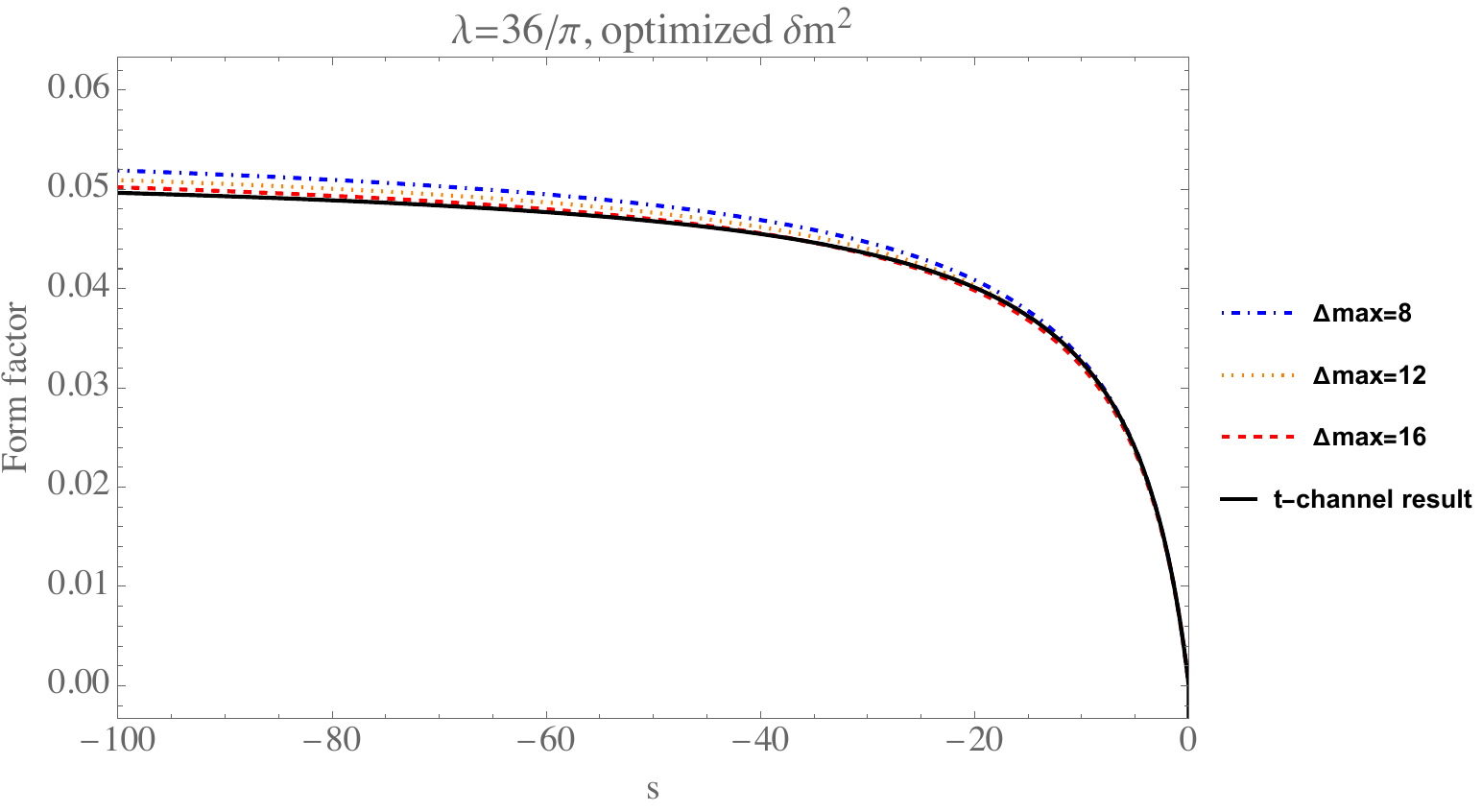}
  \caption{Strong coupling form factor comparison between the result from this paper and the `t-channel' result ({\it solid, black}) from \cite{Chen:2021bmm}, at $\lambda = \frac{6}{\pi}, \frac{36}{\pi}$. The `$\delta m^2$ from eigenvalue' plots use $\delta m^2=m_0^2-m_p^2$, with $m_p^2(\Delta_{\rm max})$ being the lowest eigenvalue of the truncated mass-squared operator $P^2$. The `optimized $\delta m^2$' plots require that the truncation result and t-channel result match at $s=-5$.}
  \label{fig:strongff}
\end{figure}

Finally, in this section we will check our results at strong coupling. The main comparison we will make is with results at $s<0$ obtained previously \cite{Chen:2021bmm} by evaluating the form factor  using the alternate expression (\ref{eq:TChannel}) that only requires single-particle eigenstates, which are easy to obtain with Hamiltonian truncation.  We compared these results at two different couplings $\lambda=6/\pi$ and $\lambda=36/\pi$. As shown in Fig.~\ref{fig:strongff}, the truncation result matches  the result from \cite{Chen:2021bmm} to about 80\% accuracy; however, we actually find that this accuracy does {\it not} seem to approach 100\%  at large $\Delta_{\rm max}$. 

\begin{figure}[t!]
  \centering
  \includegraphics[width=0.50\textwidth]{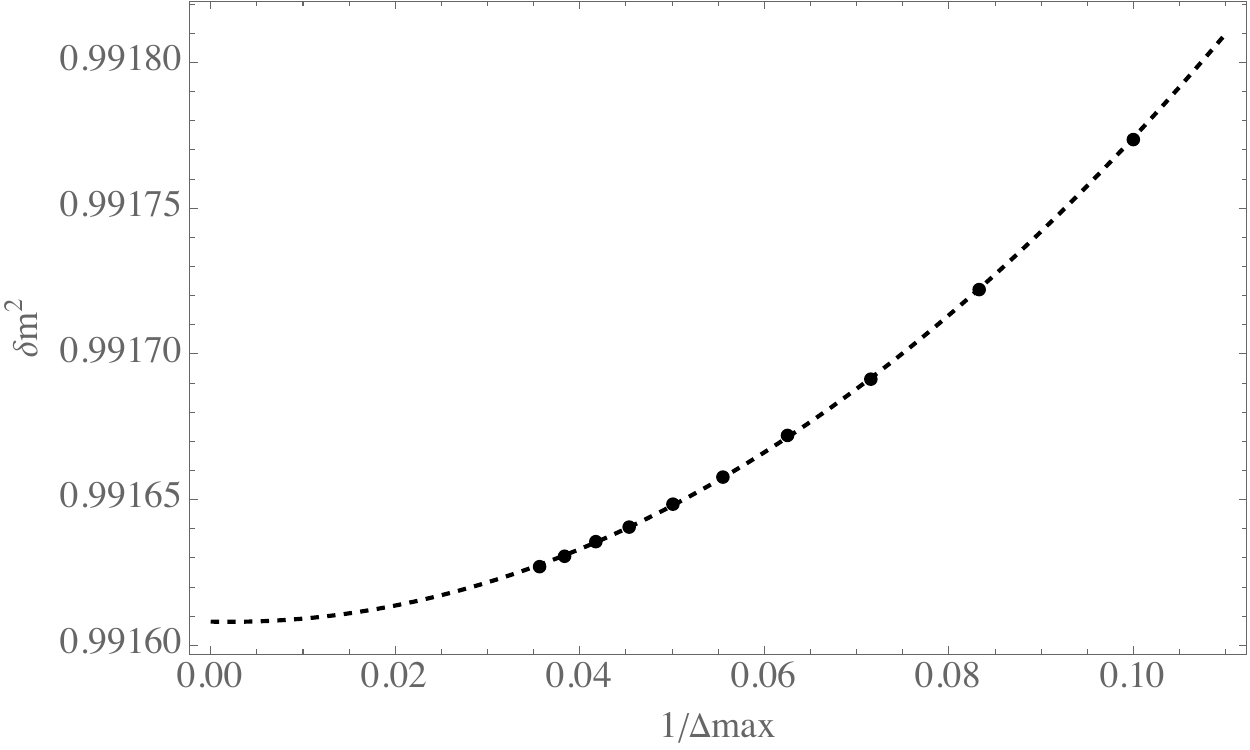}
  \includegraphics[width=0.475\textwidth]{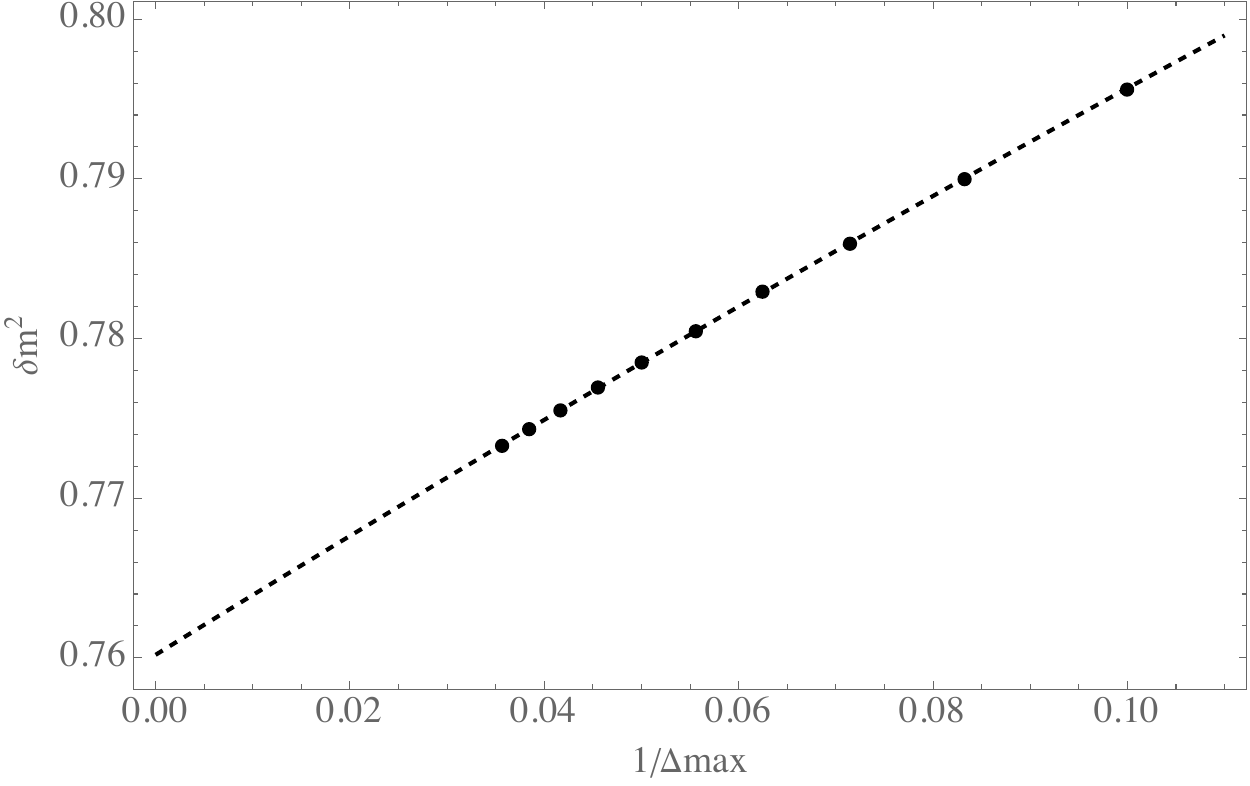}
  \caption{$\Delta_{\rm max}$ dependence of the mass shift $\delta m^2$ at weak (left) and strong (right) coupling.  The former converges quadratically like $\sim \Delta_{\rm max}^{-2}$ and the latter converges linearly like $\sim \Delta_{\rm max}^{-1}$.}
  \label{fig:dm2}
\end{figure}

What might be going wrong? 
As shown in Fig.~\ref{fig:2loopcompcoef}, we expect the individual $\phi$ and $\phi^3$ contributions to grow with $\mathcal{O}(\Delta_{\operatorname{max}})$, and therefore a slight deviation of $\delta m^2$ of order $\mathcal{O}(1/\Delta_{\operatorname{max}})$ can result in a major shift in the final form factor. We did not have to face this issue at weak coupling due to the fact that in that case the mass shift $\delta m^2$ converged sufficiently quickly, like $\sim \Delta_{\rm max}^{-2}$, to its exact value that these small truncation deviations in $\delta m^2$ did not contaminate the residual finite terms $\sim \CO(\Delta_{\rm max}^0)$ in the form factor.  However, truncation errors in  $\delta m^2$  become significant at finite coupling  because now it converges more slowly. As shown in Fig.~\ref{fig:dm2}, at small coupling, $\delta m^2$ converges quadratically, whereas at strong coupling $\delta m^2$ converges linearly. 

 More conceptually, because $\Delta_{\rm max}$ is a UV cutoff, it causes shifts in the bare parameters which can be absorbed into counterterms.  At strong coupling, these corrections are $\CO(1/\Delta_{\rm max})$, and because the $\phi$ form factor alone has a $\CO(\Delta_{\rm max})$ divergence, the $\CO(1/\Delta_{\rm max})$ corrections to $\delta m^2$ lead to {\it finite} corrections in the final form factor. 

\begin{figure}[t!]
  \centering
  \includegraphics[width=0.8\textwidth]{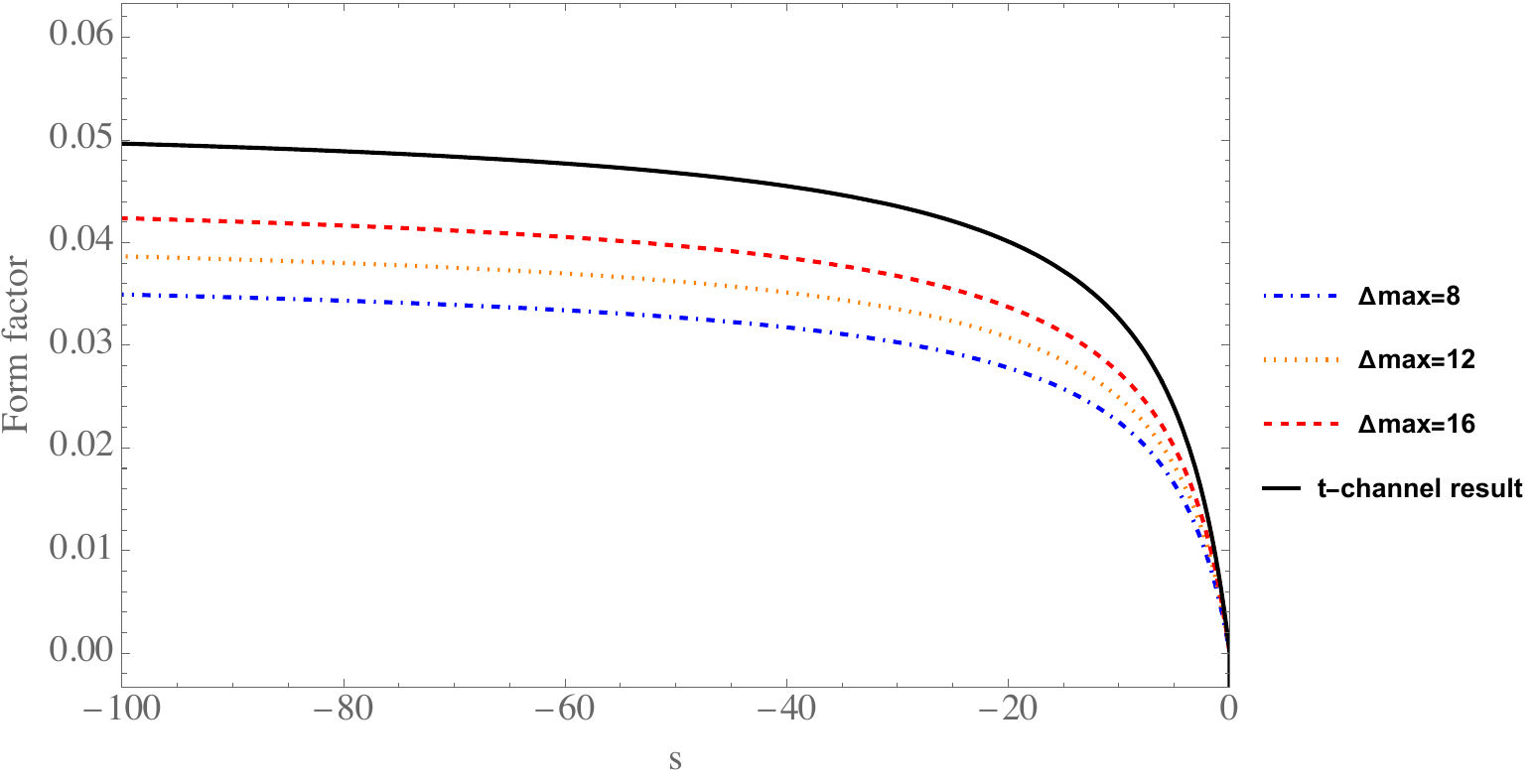}
  \caption{Strong coupling form factor comparison between truncation and t-channel results at $\lambda =\frac{36}{\pi}$, after fixing $\delta m^2=m_0^2-e_2/4$.}
  \label{fig:evenmass}
\end{figure}

There are various ways we might  attempt to  fix the value of $\delta m^2$. In our previous computations, we determined $\delta m^2$ from the eigenstate spectrum of the truncated Hamiltonian.  In particular, we obtained $\delta m^2$ by setting $m_p^2=e_1$, where $e_1$ is the lowest eigenvalue of the Hamiltonian.
 Alternatively, we could try using the two-particle eigenstate to compute $m_p^2$ by setting $m_p^2=e_2/4$, where $e_2$ is the second lowest eigenvalue of the Hamiltonian, i.e. the two-particle mass. The results are shown in Fig.~\ref{fig:evenmass}. In this plot, the truncation result appears to be converging to the t-channel result, but has a much larger error than the previous attempt and the convergence is slower than we would like.  In fact the issue with slow convergence becomes worse in this case at weak coupling.  The reason is that at weak coupling, the perturbative corrections to $e_2$ are smaller, so the truncation errors (which are present in $e_2$ even in the free theory) swamp them and therefore dominate the value of $\delta m^2$ until one reaches much larger truncations. In fact at the smaller coupling $\lambda=6/\pi$, within the range of values of $\Delta_{\rm max}$ that we used, the two-particle mass is greater than $m_0$ and $\delta m^2$ becomes negative, whereas the physical mass shift is positive.
So at best, using the second-lowest eigenvalue from truncation to set the value for $\delta m^2$ appears to converge slowly.

Finally, instead of using truncation eigenvalues, we can try to fix the value of $\delta m^2$ 
by matching the form factor result with our LSZ method to the form factor result from \cite{Chen:2021bmm} at a single value of $s$.  We choose to do this at $s=-5$,
 and the resulting form factor is plotted in green in Fig.~\ref{fig:strongff}. The fact that the two results still match within 3\% over the full range of $s$ shown, after tuning only a single parameter, is a promising sign for this approach.
 In principle, the value of $\delta m^2$ should be $\Delta_{\rm max}$-dependent but not observable dependent, and it would be good to check if the same value of $\delta m^2$ gives accurate results when used across several different observables rather than just the $T_{--}$ form factor.

\section{Future Directions}
\label{sec:future}

In this paper we have focused on the two-particle form factor $\< \Omega | T_{\mu\nu} | p_1, p_2\>$ in 2d $\phi^4$ theory, as probably the simplest possible example in which to try out the LSZ proposal from \cite{Henning:2022xlj} and explore potential issues.  The method itself is designed to be more general and hopefully it can be applied across a fairly wide range of models and multiparticle observables.  The next simplest case to study would be a two-to-two scattering amplitude in 2d $\phi^4$ theory, especially in the inelastic regime.\footnote{In fact we have learned that this is currently under investigation, \cite{MattLSZ}.}  Another important direction to explore is models where the light particle spectrum is composed of bound states, such as 2d QCD in particular, so that LSZ requires the use of composite operators which may complicate the strategy of using the equations of motion.  Even more generally, there are models (one of the simplest being 2d Ising Field Theory with a $\sigma$ deformation) without a Lagrangian description and therefore no obvious equations of motion at all.  In this case, one might still resort to the strategy mentioned in section \ref{eom} of evaluating $(\partial^2 + m^2) \CO$ by taking the  commutator $[P_+, \CO]$ with a much larger basis of states inserted as intermediate states in the matrix product.  Although we used LCT in this paper, the proposal itself is more general and it should be possible to apply it using equal-time truncation methods applied to 2d $\phi^4$ theory\cite{Rychkov:2014eea,Rychkov:2015vap}. Finally, it is important to test the method in higher dimensions than $d=2$.  The original proposal \cite{Henning:2022xlj} studied the $O(N)$ model in $d=3$ in the large $N$ limit, and a natural next step is to do 3d $\phi^4$ theory, which has been studied in LCT in \cite{Anand:2020qnp} and in $d>2$ in \cite{Elias-Miro:2020qwz,Hogervorst:2014rta,EliasMiro:2022pua,Cohen:2021erm}.

Returning to the case we studied in this paper, there are still a number of issues to investigate further.  At strong coupling, we had to choose the mass-squared shift $\delta m^2$ that appears in the LSZ prescription formula by matching the form factor to an independent calculation.  It would be good to check if this same value of $\delta m^2$ agrees across multiple observables, and to have a deeper understanding of what fixes its value in the truncation framework.  We also found that the accuracy of the form factor at a fixed  truncation parameter $\Delta_{\rm max}$ is better at small $s$ and  worse at larger $s$, and in fact if the limit $s\rightarrow \infty$ is taken before the limit $\Delta_{\rm max} \rightarrow \infty$, then the form factor apparently does not converge to the correct answer (see Fig.~\ref{fig:2loopnegs}).  It would be useful if there were a way to improve the method to obtain more accurate results at large $s$, or at least to have a better understanding of what range of $s$ is expected to be valid for a given $\Delta_{\rm max}$.

\bibliographystyle{JHEP}
\bibliography{refs}

\end{document}